\documentclass[a4paper,11pt]{article}
\usepackage{jheppub} 
\pdfoutput=1
\usepackage{graphicx}
\usepackage{makecell,booktabs,array}
\newcolumntype{x}[1]{%
>{\centering\hspace{0pt}}p{#1}}%

\usepackage[greek,english]{babel}

\newcommand{\cpi}{\text{\greektext p}}

\usepackage{tikz}
\usetikzlibrary{calc}
\usepackage{tikz-3dplot}

\usepackage{amsmath}
\usepackage{amssymb}
\usepackage{subcaption}
\usepackage{physics}
\usepackage{ulem}

\usepackage[most]{tcolorbox}

\usepackage{enumitem}

\usepackage{mathabx}  

\usepackage{graphicx}
\usepackage{dcolumn}
\usepackage{bm}
\usepackage{multirow}

\usepackage{color,soul}
\usepackage{tikz}
\usepackage{hyperref}
\usepackage[dvipsnames]{xcolor}

\newcommand{\uv}{\text{uv}}
\newcommand{\ir}{\text{ir}}

\definecolor{darkyellow}{rgb}{0.5, 0.5, 0.0}
\definecolor{darkpurple}{rgb}{0.5, 0.2, 0.8}
\definecolor{darkblue}{rgb}{0.0, 0.0, 0.8}
\definecolor{darkgreen}{rgb}{0.0, 0.4, 0.0}
\definecolor{darkred}{rgb}{0.5, 0.0, 0.0}
\usepackage{hyperref}
\hypersetup{
    linktocpage,
     colorlinks,
     citecolor=darkblue,
     linkcolor= darkblue,
     urlcolor=darkblue
}

\title{Confinement in Holographic Theories at Finite Theta}

\author{Rashmish K. Mishra}

\affiliation{Jefferson Physical Laboratory, Harvard University,\\Cambridge, MA 02138, USA}

\emailAdd{rashmishmishra@fas.harvard.edu}

\abstract{
A strongly coupled confining gauge theory with a non-zero vacuum angle undergoing a deconfinement to confinement phase transition is studied in the holographic gravitational description. A simplified five-dimensional setup is constructed where a bulk scalar models the effect of the vacuum angle, and the suitable boundary conditions on the ultra-violet (UV) and the infra-red (IR) boundaries are identified. The IR boundary condition is motivated by higher dimensional examples where the bulk scalar comes from a Wilson loop on a shrinking cycle. In this five-dimensional dual geometry, and in the limit of small backreaction in the infra-red, the critical temperature for the phase transition is shown to reduce quadratically with the vacuum angle, matching lattice results. The topological susceptibility has a sharp reduction across the critical temperature, also matching lattice results. The rate for the phase transition is estimated as a function of the vacuum angle, and is seen to be enhanced (reduced) when the field theory has a relevant (irrelevant) deformation at high energies. Crucially, for the irrelevant case, the confined phase can get destabilized for a range of parameters. In the context of early universe dynamics, if the vacuum angle is time-dependent, the transition history changes strongly: the deconfined phase can last till much lower temperatures than naively expected, and one can trigger a transition to the confined phase by a change in the vacuum angle, thus providing a controlled way to generate supercooling. As a phenomenological application, the peak frequency and the power of resulting gravitational wave signal from bubble collisions can be modified, potentially affecting their visibility in detectors. Possible generalizations of the scenario are discussed.
}

\begin{document}
\maketitle
\flushbottom

\section{Introduction}
\label{sec:Intro}
Non-perturbative phenomena and topological structure are some of the most interesting aspects of Quantum Field Theories (QFT).
In strongly coupled non-abelian gauge theories, these manifest as a dynamically generated infra-red scale (associated with confinement and mass-gap) and in the vacuum angle $\theta$. An interplay between these aspects is natural to consider. 
The phase structure of a strongly coupled confining gauge theory at finite temperature and at finite vacuum angle is a prototype of such an interplay. Naively, $\theta$ is a single dimensionless parameter in the UV, and yet we know that even at zero temperature, a rich phase structure emerges in the IR~\cite{Callan:1976je, Jackiw:1976pf, Witten:1998uka, Gaiotto:2017yup, Aitken:2018mbb}. It is natural to ask what is the effect of a non-zero temperature. Apart from theoretical importance, this question is quite relevant for phenomenology---early universe dynamics in strongly coupled dark sectors in a beyond the Standard Model (BSM) theory, which confine and have a non-zero $\theta$, can be very different than when $\theta$ is zero or when $\theta$ is slowly varying with time/temperature in the early universe. Various estimates for the resulting signals from early universe dynamics in these theories can change significantly in the presence of $\theta$.

There are not many tools to investigate these questions quantitatively. One standard approach is lattice, but simulating a theory with a non-zero $\theta$ is challenging because $\theta$ contributes as a phase to the action in Euclidean signature, ruining the manifestly positive weight of each field configuration. Another approach is to use large $N$ and holography, which allows quantitative answers, albeit it is not a guarantee that the results are general for all theories, away from when the holographic description is justified. When it is justified, the holographic approach allows going far, because a lot of structure is fixed by requirements of UV completion (i.e. from the supergravity action arising in string theory), and one can study non-equilibrium phenomena as well. One can also use these two approaches, lattice and holography, in tandem, to reinforce the advantages of each of them on the other.

The top-down inspired holographic models are often ten-dimensional, and a lot of structure is fixed in them~\cite{Bigazzi:2015bna, Bartolini:2016dbk}. But there is a lot of irrelevant structure too. It would be quite useful to have simpler 5D models that capture the right dynamics, are simple enough to handle numerical and analytical approaches, but complex enough to retain the important physics. The advantage of such a bottom-up model is that it is not tied to a specific microscopic theory, but rather captures the universal results. 

In fact, such bottom-up models have been well utilized in the literature to model properties of QCD (using lattice results and experimental results to fix the free parameters), as well as of a strongly coupled confining sector in a BSM setup. Being informed by top-down models is particularly important for BSM scenarios, because we have fewer clues there. Without it, there is a chance of missing important physics. For example, in a recent work investigating the behavior of free energy of the deconfined phase as a function of temperature (and at $\theta = 0$), a qualitatively new behavior was observed, and the underlying physics was identified~\cite{Mishra:2024ehr}. This behavior was seen already in top-down models~\cite{Buchel:2021yay} and once it was dissected, it became clear what assumptions were being made in the bottom-up constructions that were shielding it. 

In this work, we would like to understand the role of $\theta$ in the deconfinement to confinement phase transition of a strongly coupled confining theory, in a five-dimensional holographic model that is informed by top-down models. The corresponding $\theta = 0$ question has been well studied in the literature~\cite{Creminelli:2001th, Kaplan:2006yi, Hassanain:2007js, Konstandin:2010cd, Dillon:2017ctw, Bunk:2017fic, vonHarling:2017yew, Baratella:2018pxi, Fujikura:2019oyi, Agashe:2019lhy, Agashe:2020lfz, Agrawal:2021alq, Csaki:2023pwy, Eroncel:2023uqf, Mishra:2023kiu, Gherghetta:2025krk}, at least in certain limits,\footnote{E.g., when the backreaction is small in the IR, or when the deformation of the dual CFT is still somewhat weak, a lot of research has addressed questions related to the phase transition. When the backreaction is large, ref.~\cite{Mishra:2024ehr} made some progress, and there are several open questions that need further investigation.} and the order, the critical temperature, and the rate for the phase transition have been calculated. There is phenomenological interest in this question because if the phase transition is first order (as is the case), the transition proceeds through bubble collision, which generates gravitational wave signals, accessible in the frequency range to which we currently have sensitivity. The story at finite $\theta$ is less explored, and this is the gap we fill in the present work. Since even at zero temperature, $\theta$ leads to interesting IR structure, there is every reason to believe a non-trivial effect on the dynamics of the deconfinement to confinement phase transition. In the 5D models we will consider in this work, we will stay in the limit of small backreaction in the IR, which simplifies the calculations considerably. We will however pose some interesting questions that become relevant when this assumption is relaxed, and defer a more detailed analysis to a future work. 

As we will show, in the limit of small backreaction in the IR, when the theory can be studied in the so called radion effective field theory, presence of $\theta$ generates a quartic contribution to the radion potential. For a certain choice of parameters, the confining minimum gets destabilized. When this is not the case, the phase transition remains first order, and the critical temperature decreases quadratically with $\theta$. The effect is suppressed by large $N$. We will also show in the process that the topological susceptibility has a sharp drop at the critical temperature and vanishes in the deconfined phase. These results match the lattice expectations. We will discuss that this latter result is quite general and can be easily understood in the holographic context. Having established the utility of the model, we will then estimate the effect of $\theta$ on the rate of phase transition. We will then discuss the early universe scenario when $\theta$ is changing slowly, and what it means for the dynamics. 

The organization of the paper is as follows. In section~\ref{sec:Review} we review the known results about the phase transition calculation in 5D models at $\theta = 0$. Section~\ref{sec:Model} develops the 5D model, informed by top-down examples. Sections~\ref{sec:RadionPotential},~\ref{sec:CriticalTemperature},~\ref{sec:TopologicalSusceptibility} and~\ref{sec:TransitionRate} derive the radion potential at non-zero $\theta$ and use that to calculate the critical temperature, the topological susceptibility and the rate for the phase transition. Section~\ref{sec:EarlyUniverse} discusses the effect on the early universe dynamics. Section~\ref{sec:Generalizations} discusses generalizations and possible issues. Finally section~\ref{sec:Conclusion} contains the conclusions and future directions.

\section{Dynamics at Zero Vacuum Angle: A Quick Review}
\label{sec:Review}
In this section we review the calculations (and the assumptions) that exist in the literature, to answer various questions about the statics and dynamics of a deconfinement to confinement phase transition in a simplified 5D model. The starting point is the 5D Einstein-Hilbert term with a constant negative cosmological constant, two localized 3-branes with constant tension, and a scalar $\Phi$, with the combined action
\begin{align}
    S =  \int \dd^5 x \sqrt{g} \left(2 M_5^3\mathcal{R} - \Lambda\right) + \int \dd^5 x \sqrt{g} \left(-\frac12 (\partial\Phi)^2 - V(\Phi)\right) + \sum_{i = \uv, \ir} \int \dd^4 x \sqrt{g_i} \, T_i\:.
\end{align}
An exact time independent solution to the resulting Einstein equations is a slice of 5D Anti-de Sitter (AdS) space, truncated by the two branes. In the Fefferman-Graham coordinate $z$, the metric is given by
\begin{align}
    \dd s^2 = \frac{\ell^2}{z^2}\left(-\dd t^2 + \dd \vec{x}^2 + \dd z^2\right)\:,
    \label{eq:metric-confined}
\end{align}
where $\ell$ is the AdS radius, and the asymptotic AdS boundary is at $z = 0$. The bulk cosmological constant and the brane tensions are constrained to satisfy 
\begin{align}
    \Lambda/M_5^3 = -24/\ell^2\:, \:\: T_\uv/M_5^3 = -T_\ir/M_5^3 = 24/\ell\:.
    \label{eq:brane-tensions-and-bulk-cosmological-constant}
\end{align}
The brane locations can be chosen to be at $z = z_\uv$ (UV brane) and $z = z_\ir$ (IR brane), so that the spacetime is a slice of the full AdS space. The scalar $\Phi$, with its bulk potential $V(\Phi)$, and its boundary conditions on the UV/IR branes (engineered by brane localized potentials) develops a background value that varies along the bulk $z$ direction. For a given choice of parameters, the action is minimized for a specific value of $z_\ir$, which was a free parameter (therefore a modulus in the KK reduced 4D theory) before. Without much tuning, one can easily engineer $z_\ir \gg z_\uv$. This can be done both when the backreaction from the scalar is small in the IR~\cite{Goldberger:1999uk}, or when it is appreciable~\cite{DeWolfe:1999cp}.

This solution is the celebrated Randall-Sundrum (RS) geometry~\cite{Randall:1999ee}, with $\Phi$ the Goldberger-Wise (GW) scalar~\cite{Goldberger:1999uk}. Holographically, the setup describes a strongly coupled confining 4D CFT that couples to 4D gravity and has a deformation in the UV~\cite{Rattazzi:2000hs, Arkani-Hamed:2000ijo}. The coupling to gravity comes from requiring a UV brane that makes the 5D volume finite, confinement comes from the spacetime ending in the IR at a fixed location so that the string worldsheet dual to probe Wilson loops are bounded, and the deformation comes from the bulk potential for $\Phi$ because the potential is dual to the beta-function for a scalar deformation of the CFT.

A second solution to the Einstein equations exists, and is the AdS-Schwarzschild (AdSS) solution, with the metric
\begin{align}
    \dd s^2 = \frac{\ell^2}{z^2}\left(-\left(1-\frac{z^4}{z_h^4}\right)\dd t^2 + \dd \vec{x}^2 + \left(1-\frac{z^4}{z_h^4}\right)^{-1} \dd z^2\right)\:,
    \label{eq:metric-deconfined}
\end{align}
and a UV brane at $z = z_\uv$, with the same tension as in eq.~\eqref{eq:brane-tensions-and-bulk-cosmological-constant}, but crucially, no IR brane in the bulk. Rather, there is a planar horizon localized in the bulk at $z = z_h$, extended in the $\mathbb{R}^3$ direction. This is the black brane solution. Crucially, this is a \textit{time-dependent}  solution, and the time-dependence comes from a mismatch in the junction conditions at $z = z_\uv$, which makes the horizon move to larger values of $z$. In the limit of $z_\uv\to0$, i.e. the UV brane sent to the boundary, this becomes a time-independent solution. 

This second solution is dual to the deconfined phase of the CFT.\footnote{This can be seen easily by the infinite red-shift of the tension of the string worldsheet anchored to the Wilson loop on the boundary, since $g_{tt} = 0$ at the horizon.} The temperature $T$ of the CFT can be identified by the asymptotic value of the periodicity of time $t$ after continuing it to Euclidean signature, and requiring a smooth manifold. The standard result is that 
\begin{align}
    T = \frac{1}{\cpi\,z_h}\:.
    \label{eq:temperature-vs-zh}
\end{align}
The time dependence of the solution results in $z_h \to z_h(t)$, or effectively, $T\to T(t)$. In the limit of the UV brane sent to the boundary, i.e. making the 4D gravity decouple, one recovers the time-independent solution, which is the pure AdS-Schwarzschild solution.\footnote{One quick way to understand the time-dependence is to note that in the dual CFT at finite temperature and coupled to gravity, the energy density $\propto \,  T^4$ redshifts under time-evolution, which is equivalent to increasing Planck scale, and therefore to the UV brane moving to the boundary. Choosing coordinates so that the UV brane is fixed, the black brane moves to larger values of $z_h$, which is consistent with a decreasing temperature by eq.~\eqref{eq:temperature-vs-zh}.} 

The statics and dynamics of the phase transition have been studied in the limit of small backreaction in the IR, i.e. to say that the effect of $\Phi$ is not enough to deform the geometry near the IR brane and near the horizon, away from the pure AdS and pure AdSS limits respectively. This is of course an idealization, not often seen in UV complete examples, and there are important consequences when this is relaxed~\cite{Mishra:2024ehr}. We will however stay in the limit of small backreaction in this work, but will comment on relaxing this assumption in sec.~\ref{sec:Generalizations}. In this small backreaction limit, the 5D theory can be well described after Kaluza-Klein (KK) reduction, by a 4D EFT. The light degrees of freedom in this EFT are the 4D graviton and a scalar, the radion, which we will denote by $\varphi$. The cutoff of this EFT is set by the mass of the KK gravitons $m_\text{KK}$. The mass of the radion $\varphi$, relative to $m_\text{KK}$ is a measure of the backreaction in the IR and since we are staying in the limit of small backreaction, this ratio is small, hence justifying the EFT.

In the radion EFT, the action is given as
\begin{align}
    S = \int \dd^4 x \sqrt{g} \left(
    2 M_5^3\ell\,R \left(1-\varphi^2\right) - 12 M_5^3\ell\,(\partial\varphi)^2 - V(\varphi) + V_0
    \right)\:.
    \label{eq:4d-action-gravity-and-radion}
\end{align}
In our normalization, the radion $\varphi$ is dimensionless. Here $V_0$ is a constant, chosen by hand to ensure the 4D cosmological constant is zero.\footnote{One can wonder why does one have to do this by hand, if the 5D geometry is such that for each value of $z$ the 4D geometry is flat. The answer is because the procedure of obtaining the EFT does not exactly solve the Einstein equations. A more principled approach arrives at the same conclusions~\cite{Lust:2025vyz}.}
Also, the term $\varphi^2 R$ can be ignored if we are interested in flat space, for which $R=0$ on-shell. 
The potential $V(\varphi)$ in general receives contributions both from the bulk potential for $\Phi$ as well as the brane localized potentials for $\Phi$. Without $\Phi$, the potential for the radion vanishes identically, enforced by the relations among $\Lambda, T_\uv, T_\ir$ in eq.~\eqref{eq:brane-tensions-and-bulk-cosmological-constant}. These conditions, resulting from the requirement of solving the 5D Einstein equations, ensure that $\varphi$ is a modulus and that the 4D cosmological constant is zero. Adding a $\Phi$ sector is motivated by generating a natural expectation value for the radion $\varphi$, but the 4D cosmological constant still has to be tuned away. Therefore, with a $\Phi$ sector, one can slightly modify the relations in eq.~\eqref{eq:brane-tensions-and-bulk-cosmological-constant} as long as one enforces the resulting radion potential to vanish at the minimum. 

Given the radion potential, the free energy (per unit 3-volume) $f$ of the two phases is straightforward to obtain. The first step is to obtain the free energy when there is no $\Phi$ and therefore no radion potential. The UV brane is also sent to the boundary to make the discussions simpler. In this limit, the free energy of both the phases is formally divergent, but the difference is finite, and is given as~\cite{Creminelli:2001th}
\begin{align}
    f_\text{d} - f_\text{c} = -2\cpi^4 (M_5\ell)^3 T^4\:.
\end{align}
In this limit, the deconfined phase always has a lower free energy and is always the preferred phase.\footnote{Without a $\Phi$, there is no deformation of the CFT and no other scale apart from the temperature.} Once a $\Phi$ sector is added and the radion potential $V(\varphi)$ is generated, the difference in free energies is modified to
\begin{align}
    f_\text{d} - f_\text{c} = -2\cpi^4 (M_5\ell)^3 T^4 + V_0\:,
\end{align}
where $V_0$ is fixed by the requirement of a vanishing 4D cosmological constant. 

The critical temperature $T_c$, defined as the temperature when the free energy of the two phases is equal, is given as
\begin{align}
    T_c = \left(\frac{-V(\left<\varphi\right>)}{2\cpi^4 (M_5\ell)^3}\right)^{1/4}=\left(\frac{V_0}{2\cpi^4 (M_5\ell)^3}\right)^{1/4}\:.
    \label{eq:critical-temperature-no-theta}
\end{align}
For temperatures below $T_c$, the deconfined phase has a higher free energy, and is metastable. All these results are in the limit of the UV brane sent to the boundary, and are slightly corrected when the UV brane is brought back to a finite location (these corrections are suppressed by the 4D Planck scale).

For temperatures below $T_c$, the metastable deconfined phase is expected to transition to the more stable confined phase. While it does not, the metastable phase has a positive cosmological constant (since it is metastable, and the true vacuum has zero cosmological constant, by construction) and inflates if the 4D Planck scale is finite. The rate for the transition process is subtle to calculate, and requires making several further approximations which while allowing a quantitative expression, might only be of limited validity. Some of these aspects are currently under active scrutiny. We will here content ourselves with the agreed results for the rate. The rate proceeds through a bubble of true vacuum nucleating inside the false vacuum, and the standard results from vacuum tunneling at finite temperature can be utilized. At finite temperatures, the $O(3)$ bounce dominates,\footnote{See ref.~\cite{Gherghetta:2025krk} for a more accurate estimate for the relevant bounce which interpolates between $O(3)$ and $O(4)$ solutions.} and the rate (per unit 3-volume) for the phase transition is given as
\begin{align}
    \Gamma = T^4 \exp\left(-a\frac{(M_5\ell)^3}{\delta}f(T/T_c)\right)\:,
\end{align}
where $a$ is an order 1 numerical factor, $\delta$ is a parameter that depends on the parameters in the radion potential and is a measure of explicit breaking of CFT in the IR, and $f(T/T_c)$ encodes the temperature dependence, being infinite at $T=T_c$ (so that the rate is zero there), and monotonically decreases as $T$ gets smaller than $T_c$. Various factors in this expression can be understood in a straightforward way. The dependence on $(M_5\ell)^3 \sim N^2$ is expected because the rate to go from a phase with $N^2$ degrees of freedom to order one degrees of freedom will be suppressed by $e^{-N^2}$. If the explicit breaking of CFT is zero, we know that there is no special temperature and the rate should vanish, which is consistent with $e^{-1/\delta}$ dependence. The rate should only be non-zero at temperatures below the critical temperature $T_c$, hence there should be functional dependence on $T/T_c$ to ensure this. Finally, the multiplicative factor of $T^4$ is expected on dimensional grounds.  

If we now bring back the UV brane to a finite location (i.e. the 4D Planck scale is large but finite), the metastable phase starts to inflate, since it has a positive vacuum energy. As a heuristic for the phase transition to complete, the rate in a Hubble 4-volume should be order 1 or larger, 
\begin{align}
    \Gamma/H^4 \gtrsim 1\:,
\end{align}
which can be used to solve for the temperature at which this condition is satisfied, the nucleation temperature $T_n$. For larger values of $M_5\ell$, the nucleation temperature $T_n$ is smaller (than the critical temperature $T_c$), and this generates a supercooled state during which the deconfined phase, while being metastable, gets to live. Note that a large $M_5\ell$ is necessary for theoretical control of the 5D gravitational theory, which generates a tension between a successful transition to the stable phase and a trustworthy regime of validity of the calculations leading up to it. Finding a scenario that is under theoretical control and has a successful transition to the confined phase has driven a lot of research in this field, see refs.~\cite{vonHarling:2017yew, Baratella:2018pxi, Agashe:2019lhy, Agashe:2020lfz, Csaki:2023pwy, Mishra:2023kiu}.

This completes a very fast review of the last quarter century worth of developments in this direction. The reader is referred to the references cited earlier for further details.

\section{Simplified Model for a Non-zero Vacuum Angle}
\label{sec:Model}
In this section, we present a simplified model in a 5D AdS background that captures the underlying physics of a finite $\theta$. For a $SU(N)$ gauge theory on $\mathbb{R}^{3,1}$, the action at finite $\theta$ is given as
\begin{align}
    S = \int \dd^4 x \,\, \tr \left(-\frac{1}{2}F^2 + \frac{\theta\,g^2}{8\cpi^2}F\widetilde{F}\right)\:,
    \label{eq:4d-action-with-theta}
\end{align}
where $F$ is the field strength, $g$ is the gauge coupling of the theory, and the vacuum angle $\theta$ is periodic with period $2\cpi$. The resulting dynamics is expected to be invariant under $\theta \to \theta + 2\cpi$. 

In the large $N$ limit, for fixed `t Hooft coupling $\lambda_t = g^2N$, the action is given as
\begin{align}
    S = \int \dd^4 x \,\, \tr \left(-\frac{1}{2}F^2 + \frac{\lambda_t}{8\cpi^2}\,\frac{\theta}{N}\,F\widetilde{F}\right)\:,
\end{align}
which makes it manifest that all effects of $\theta$ come with a $1/N$ factor. In fact, in the $N\to\infty$ limit, naively, the effect of $\theta$ disappears. As is well known, some effects can survive in this limit, if they are inherently $N$ enhanced. For example, the vacuum energy of the theory at finite $\theta$, $E(\theta)$ is famously known to scale as~\cite{Witten:1998uka}
\begin{align}
    E(\theta) = N^2 \, \min_k \, h((\theta + 2\cpi k)/N) = \, \min_k \, C \left(\theta + 2\cpi k\right)^2 + \mathcal{O}(1/N)\:,
\end{align}
where $k\in \mathbb{Z}$ is an integer, $h(\theta)$ is some unknown function of $\theta$, and $C$ is an $N$ independent constant.

How should one model $\theta$ in a bottom-up gravitational setup? The first observation can be that apart from a stress energy tensor that the gauge theory must have (being a local QFT), there are two local gauge invariant scalar operators $\tr F^2$ and $\tr F\widetilde{F}$ with low enough scaling dimensions. They are dual to scalar fields in the 5D dual, using standard AdS/CFT dictionary, and must be included in the action. A non-zero $\theta$ acts as a source for the operator $\tr F\widetilde{F}$, and must be included as such in the 5D picture for the dual scalar. Apart from these two scalars, we must have the 5D gravity. Therefore, a putative bottom-up model can be
\begin{align}
    S_5 = \int \dd^5 x \sqrt{g}
    \left(
    2 M_5^3 \mathcal{R} - \Lambda
    -\frac{1}{2}(\partial\Phi)^2 - \frac{1}{2}f_\sigma^3(\Phi)(\partial\sigma)^2 - V(\Phi, \sigma)
    \right)\:,
    \label{eq:5D-action-with-gravity-and-two-scalars}
\end{align}
where $M_5$ is the 5D Planck scale, $\mathcal{R}$ is the Ricci scalar, $\Lambda$ is the 5D (negative) cosmological constant, $\Phi$ is the field dual to $\tr F^2$, $\sigma$ is the field dual to $\tr F\widetilde{F}$ and is periodic with period $2\cpi$. To be general, we have considered the decay constant $f_\sigma$ to be $\Phi$ dependent for now, to account for the $g^2$ factor in the second term in eq.~\eqref{eq:4d-action-with-theta}.\footnote{The rescaling $F\to g\, F$, which removes the $g^2$ factor multiplying $\theta$, corresponds to a field redefinition in the dual theory.}

The potential term $V(\Phi, \sigma)$ models the running of the operators in the field theory. Being a periodic variable, $\theta$ does not run under the renormalization group, and we can take $V(\Phi,\sigma) = V(\Phi)$, to be $\sigma$ independent. While non-perturbative instanton effects generate terms like $\exp(-8\cpi^2/g^2)\exp(\pm i\theta)$, they are exponentially suppressed in the large $N$, fixed $\lambda$ limit. We will ignore these contributions and stick to keeping the bulk potential independent of $\sigma$. It is less clear what to take for $f_\sigma(\Phi)$. In fact, as already argued in refs.~\cite{Gursoy:2007cb, Gursoy:2007er} (see also ref.~\cite{Katz:2007tf}), one can take it to be a constant, to leading order. This certainly seems to fit well to data~\cite{Gursoy:2009jd}. We will assume that $f_\sigma$ is independent of $\Phi$ for now, and comment on including $\Phi$ dependence later. Finally, there are several choices for $V(\Phi)$. We will take it to be a simple quadratic
\begin{align}
    V(\Phi) = 2 (\epsilon/\ell^2) \Phi^2\:,
    \label{eq:GW-bulk-potential}
\end{align}
where the dimensionless quantity $\epsilon$ can be either positive or negative, and corresponds to an irrelevant/relevant deformation of the field theory in the UV. From the field theory point of view, it is more natural to consider a relevant deformation (an explicit breaking of CFT) that triggers confinement (a spontaneous breaking of CFT) at some scale, but to be general we will stay agnostic to the sign of $\epsilon$.\footnote{A significant amount of literature on RS models has considered $\epsilon > 0$, with the assumption that even though the effect of the irrelevant deformation gets progressively weaker in the IR, it can still trigger confinement.} This field $\Phi$ can be identified with the Goldberger-Wise (GW) scalar that is often used in such bottom-up models to stabilize the geometry and generate an exponentially suppressed confinement scale relative to the UV scale. The choice of a mass term only for $\Phi$ is then no surprise, although see refs.~\cite{Agashe:2020lfz, Mishra:2023kiu} that consider higher order terms in the potential to model related dynamics. We will stay with the simplest choice here, however, the steps are easy to generalize.

As stated earlier, we will work in the limit of small backreaction. We can therefore ignore the effect of $\Phi, \sigma$ on the geometry, and take the geometry to be coming from solving the pure Einstein equations. Since we want to study the confined and deconfined phases of the theory, the dual geometries to consider are the pure AdS solution between a UV and an IR brane (dual to the confined phase), and the AdS Schwarzschild solution with a UV brane and a horizon in the bulk (dual to the deconfined phase). While the presence of the UV brane is necessary to have dynamical gravity in the 4D theory that couples to the gauge theory (and is relevant when talking about the cosmology, in sec~\ref{sec:EarlyUniverse}), in the present discussion we can safely send it to the asymptotic AdS boundary without affecting the conclusions. 

The problem is therefore specified: we need to solve for the profiles of $\Phi$ and $\sigma$ in the two backgrounds, and draw conclusions about the physics from them. However, to proceed we need to decide what are the appropriate boundary conditions to impose on the two fields. 

Since $\Phi$ can be identified with the GW field, we can use the standard options often used in the literature---either a Dirichlet boundary condition on both the UV and IR brane, or a Dirichlet boundary condition on the UV and a Neumann boundary condition on the IR.\footnote{It is more natural to consider a Dirichlet boundary condition on the UV since it corresponds to switching on a deformation. It is less clear what to make of the IR boundary condition, hence a democratic choice. In a fully UV complete theory, the IR boundary conditions can be determined by other considerations, e.g. see~\cite{Brummer:2005sh}.} Note that there is no boundary condition to be imposed in the IR for the AdS Schwarzschild solution---regularity of the solution at the horizon takes care of that.  

What about the boundary conditions on the $\sigma$ field? The UV boundary condition is easy to see, and is exactly the vacuum angle of the theory that we have been trying to model. This still leaves the IR boundary condition to be specified (again note that we do not need to worry about the IR boundary condition when there is a horizon in the bulk).

At this point, one can go the same route as $\Phi$ and decide to choose either a Dirichlet or a Neumann boundary condition for $\sigma$. However it turns out one can do better. For this, we need to take a detour and understand how $\theta$ or $\sigma$ arise in more UV complete theories, with more than 5 dimensions. We will find that at least for a class of UV completions, a certain choice of boundary conditions for $\sigma$ can be motivated.

\subsection{A ten-dimensional example}
\label{subsec:Model-10D}
Let us first review a particularly useful example~\cite{Witten:1998zw, Witten:1998uka} that allows getting to the point quickly. Consider type IIA superstring theory on the manifold $\mathbb{R}^4\times\mathbb{S}^1\times\mathbb{R}^5$ with $N$ parallel $D4$ branes localized at a point in the $\mathbb{R}^5$. The spin structure is chosen such that the fermions change sign when going around the $\mathbb{S}^1$, thus breaking supersymmetry. The theory on the branes at low energies is pure $U(N)$ gauge theory in 4D and at large $N$, the $SU(N)$ part can be studied in the weakly coupled string theory on the supergravity solution generated by the branes. The generated supergravity solution has the topology of $\mathbb{R}^4\times\mathbb{D}\times\mathbb{S}^4$, where $\mathbb{D}$ is a 2D disk anchored on $\mathbb{S}^1$. The metric for the supergravity solution is 
\begin{align}
    \dd s_{10}^2 = \frac{8\cpi}{3}\eta \lambda^3 (\dd t^2 + \dd \vec{x}^2) + \frac{8}{27}\eta\lambda\cpi(\lambda^2 - \lambda^{-4})\dd \psi^2 + \frac{8\cpi}{3}\eta\lambda \frac{\dd \lambda^2}{\lambda^2 - \lambda^{-4}} + \frac{2\cpi}{3}\eta\lambda \dd\Omega_4^2\:,
\end{align}
where $(t, \vec{x})$ are coordinates on $\mathbb{R}^4$, $\Omega_4$ are coordinates on the $\mathbb{S}^4$, $(\lambda, \psi)$ are coordinates on the disk $\mathbb{D}$, with center at $\lambda = 1$ and the boundary at $\lambda\to\infty$, and $\eta \gg 1$ is a parameter that is needed for the string theory to be approximated by its supergravity limit. 

To include a vacuum angle, one adds a $U(1)$ gauge field $a$ with field strength $f_{ij}$ that arises from the Ramond-Ramond sector. The low energy theory on the world branes has a term
\begin{align}
    \frac{1}{8\cpi^2}\int a \wedge \, \tr (F\wedge F)\:,
    \label{eq:Chern-Simons-term}
\end{align}
where $F$ is the field strength of the $SU(N)$ gauge theory on the brane. The gauge invariant operators in the theory are Wilson loops and field strengths, and we can choose to switch on background values for them.  If we now require
\begin{align}
    f_{ij} = 0\:, \:\: \int_{\mathbb{S}^1}a \:\: =\:\:  \theta + 2 \cpi k\:,\: k \in \mathbb{Z}\:,
\end{align}
the 4D theory generates a $\theta$ term.  
In the supergravity dual, we have to solve the equations to see the behavior of $f$ away from the D4 branes. By AdS/CFT, the solution for $f$ must respect the conditions
\begin{align}
    \lim_{\lambda\to\infty} f_{ij} = 0\:, \:\: \lim_{\lambda\to\infty} \int_{\mathbb{S}^1}a = \theta + 2 \cpi k\:.
    \label{eq:bc-on-f-and-a}
\end{align}
Solution to the equations of motion for $f$, subject to the first condition in eq.~\eqref{eq:bc-on-f-and-a}, give the only non-zero component to be 
\begin{align}
    f_{\lambda\psi} = 6\left(\frac{\theta}{2\cpi} + k\right)\lambda^{-7}\:.
\end{align}
In the gauge $a_\lambda = 0$, we can solve for $a_\psi$ to be
\begin{align}
    a_\psi = a_0 - \left(\frac{\theta}{2\cpi} + k\right)\lambda^{-6}\:,
\end{align}
where $a_0$ is a constant. Requiring the condition on $a$ from eq.~\eqref{eq:bc-on-f-and-a} we get
\begin{align}
    a_\psi(\lambda) = \left(\frac{\theta}{2\cpi} + k\right)\left(1-\lambda^{-6}\right)\:.
\end{align}
From this solution we can define a $\lambda$ dependent angle $\Theta$ as
\begin{align}
    \Theta(\lambda) = \int_{\mathbb{S}^1}a = \left(\theta + 2 \cpi k\right)(1-\lambda^{-6})\:
\end{align}
which has the properties
\begin{align}
    \lim_{\lambda\to\infty} \Theta(\lambda) = \theta + 2\cpi k\:,\:\:
    \lim_{\lambda\to 1} \Theta(\lambda) = 0\:.
\end{align}
The second condition is not surprising: since $\lambda \to 1$ is the center of a 2D disk, an angular variable like $\Theta(\lambda)$ should vanish there. 

With this discussion, it is clear what the boundary conditions for the field $\sigma$ should be. The background value for the field $\sigma$ is to be identified with $\Theta$, as a function of the holographic direction (the one which stops at some point in the IR, $\lambda$ in the example above, and $z$ in the usual 5D AdS background). As motivated earlier, the UV boundary condition should set $\sigma$ to equal $\theta$ angle itself, which we see here explicitly. The non-trivial result is that $\sigma$ should vanish in the IR. Without this exercise, we would have had an arbitrariness in the choice of the boundary condition in the IR.\footnote{I thank Matt Reece for discussions related to this. During the completion of this work, I attended a talk by Csaba Csaki~\cite{Csaki:talk} which also advocates for the IR boundary condition motivated here.} 

Therefore, the boundary conditions to be satisfied by $\sigma$ are 
\begin{align}
    \left. \sigma \right|_\uv = \theta + 2 \cpi k\:,\:\: \left. \sigma \right|_\ir = 0\:,\:\: k\in \mathbb{Z}\: . 
    \label{eq:sigma-bc}
\end{align}

The result is actually quite general. In many supergravity solutions that are dual to gauge theories that show a mass gap and confinement, the geometry smoothly caps off at some point in the IR. Analogous to the discussion here, one can define the corresponding $\Theta$ in those examples, and one would get to the same conclusion: since it is an angular variable, it must vanish at the origin of coordinates. Alternatively, it is because the proper length of the utilized cycle vanishes in the IR. This discussion however assumes that $\theta$ comes from wrapping over a compact dimension that shrinks to zero size in the IR.\footnote{In a geometry where there are multiple dimensions that compete to shrink, if $\theta$ comes from a non-shrinking dimension, it is less clear how to uniquely fix the boundary condition.}

While we have fixed the boundary condition on $\sigma$ in our 5D simplified model, we would like to see explicitly how the form of the action we have assumed comes about. For this, the type IIA example above is not useful, because it is not asymptotic to AdS$_5 \times \mathbb{X}$ in the UV, for some smooth manifold $\mathbb{X}$. The important feature in the type IIA example was that the supergravity background had a disk and there was a gauge field wrapping around it. 

\subsection{A six-dimensional example}
\label{subsec:Model-6D}
Instead of a full 10D example, let us start with a 6D spacetime which has a disk, and is asymptotically of the form AdS$_5 \times \mathbb{X}$ in the UV. Let us consider $\mathbb{X} = \mathbb{S}^1$, for simplicity. The metric for this 6D spacetime is 
\begin{align}
    \dd s_6^2 = \frac{\ell^2}{z^2}\left(-\dd t^2 + \dd \vec{x}^2 + \dd z^2\right) + h(z)\dd y^2\:,
\end{align}
where $\ell$ is the AdS$_5$ radius, $y$ is the coordinate on the $\mathbb{S}^1$ of radius $R$, $y\cong y+ 2\cpi R$, and $h(z)$ is some unspecified function. We will assume that this geometry can be obtained from Einstein equations for some choice of matter fields. 

Since we want the geometry to be asymptotically AdS$_5\times\mathbb{S}^1$ in the UV, and that the $\mathbb{S}^1$ shrinks to zero size at some point $z = z_\ir$, we require
\begin{align}
    \lim_{z\to 0 }\: h(z) = 1\:\:, \lim_{z\to z_\ir }\: h(z) = 0\:\: .  
    \label{eq:conditions-on-f}
\end{align}
The ``disk'' is provided by the $(z, y)$ subspace, with the boundary of the disk at $z = 0$ and the center at $z = z_\ir$.

The radius $R$ is determined by the requirement of a smooth geometry near $z = z_\ir$. To see this, we can expand the metric near $z = z_\ir$. Substituting $z = z_\ir + \delta$, the metric in the $(z, y)$ subspace, to leading order in $\delta$, is 
\begin{align}
    \dd s^2 \to \frac{\ell^2}{z_\ir^2}\dd \delta ^2 + \left(\delta h'(z_\ir) + \frac12\delta^2 h''(z_\ir) + \cdots\right)\dd y^2\:,
\end{align}
where the first term in the Taylor expansion multiplying $\dd y^2$, $h(z_\ir)$, is zero by assumption. For smoothness, the metric should be of the form of polar coordinates, which immediately tells us that we additionally need $h'(z_\ir) = 0$ and $h''(z_\ir)>0$. Requiring that, the radius $R$ is fixed to be
\begin{align}
    R = \frac{\ell}{z_\ir}\sqrt{\frac{2}{h''(z_\ir)}}\:.
    \label{eq:R-for-f}
\end{align}

We now add a $U(1)$ gauge field $A$ to the theory. We have two kinds of gauge invariant operators, the field strength and the Wilson loops. As before, we switch on a zero field strength, but a non-zero winding around the $\mathbb{S}^1$ (both asymptotically, as $z\to0$)
\begin{align}
    \lim_{z \to 0}\: F_{ij} = 0\:,\:\: \lim_{z \to 0}\: \int_{\mathbb{S}^1} A = \lim_{z \to 0}\: \int \dd y A_6 = \theta + 2\cpi k\:, \:\: k \in \mathbb{Z}\:.
\end{align}
One can do the same exercise as in the previous section to solve for $F$ in the bulk and define a $z$ dependent $\Theta$. From the previous discussion we know that $\Theta$ satisfies the boundary conditions 
\begin{align}
    \Theta(0) = \theta + 2\cpi k\:,\:\: k \in \mathbb{Z}\:,\:\: \Theta(z_\ir) = 0\:.
    \label{eq:Theta-bc}
\end{align}
The goal here is to see what kind of 5D action would give a background solution $\sigma(z) = \Theta(z)$.

We start with the action
\begin{align}
    S_6 = -\frac{1}{4e_6^2}\int \dd^6 x \sqrt{g_6} F_{AB} F_{CD} g_6^{AC}g_6^{BD}\:,
\end{align}
where $e_6$ is the 6D gauge coupling with mass dimension $-1$, and the subscript $6$ on the metric denotes that it is the 6D metric. 
The equation of motion for $A_6$, in the axial gauge $A_5 = 0$, is given as
\begin{align}
    \frac{\dd}{\dd z}\left(\frac{\ell^3}{z^3}\,\frac{1}{\sqrt{h}}\,\frac{\dd A_6}{\dd z}\right) = 0\:,
\end{align}
which can be readily solved to give
\begin{align}
    A_6(z) = a_0\int \dd z \sqrt{h} \left(\frac{z}{\ell}\right)^3 + a_1\:,
\end{align}
where $a_0,a_1$ are constants. As before, we define 
\begin{align}
    \Theta(z) = \int \dd y A_6 = 2\cpi R\,A_6(z)\:, 
\end{align}
and require it to satisfy the boundary conditions in eq.~\eqref{eq:Theta-bc}. Given an $h(z)$, this will uniquely fix $F_{z6}(z)$ and $\Theta(z)$. We then identify $\Theta$ as the background value of $\sigma$.

To calculate the effective 5D action, we start with the 6D action for the gauge field $A$, substitute $\sigma = 2\cpi R A_6 $, and do the trivial $y$ integration (since nothing depends on $y$) to get
\begin{align}
    S_6 &\supset -\frac{1}{2e_6^2} \int \dd^6 x \sqrt{g_6} (\partial_A A_6)(\partial_C A_6) g_6^{AC} g_6^{66} 
    \nonumber \\
    &= -\frac{1}{4 e_6^2 \cpi R}\int \dd^5 x \sqrt{g} \frac{1}{\sqrt{h}} (\partial_A \sigma)(\partial_C \sigma) g^{AC}
    \equiv -\frac12 \int \dd^5 x \sqrt{g} \,f_\sigma^3\, h^{-1/2}\,(\partial\sigma)^2
    \:,
\end{align}
where $g$ refers to the 5D part of the metric (which is AdS$_5$). Note that in the UV ($z\to0$) $h\to 1$, and the action is that of a free scalar. The decay constant $f_\sigma$ can be identified with
\begin{align}
    f_\sigma^3 = \frac{1}{2\cpi e_6^2 R}\:.
\end{align}
Deep in the IR however, $h$ deviates from $1$, and this can be captured by making $f_\sigma$ be $\Phi$ dependent.

As an explicit example, consistent with the requirements in eq.~\eqref{eq:conditions-on-f}, we can choose a simple form for $h(z)$:
\begin{align}
    h(z) = \left(1-\frac{z}{z_\ir}\right)^2\:.
\end{align}
For this choice of $h$, eq.~\eqref{eq:R-for-f} fixes $R = \ell$, and we get 
\begin{align}
    S_5 &= -\frac12 \int \dd^5 x \sqrt{g} \, f_\sigma^3 (\partial\sigma)^2\,\left(1 - z/z_\ir\right)^{-1}\:\:,\:\: f_\sigma^3 = \frac{1}{2\cpi e_6^2\ell} \sim \ell^{-3}\:,
    \\
    \sigma(z) &= \Theta(z) = 2\cpi \ell A_6(z) = \left(\theta + 2\cpi k\right)\left(1- 5\frac{z^4}{z_\ir^4} + 4\frac{z^5}{z_\ir^5}\right)\:,
    \\
    F_{56} &= \frac{10}{\cpi \ell} (\theta+2\cpi k)\frac{z^3}{z_\ir^4}\left(\frac{z}{z_\ir}-1\right)\:.
    \label{eq:model-6D-action-and-profiles}
\end{align}
We see that the solutions have the right asymptotic forms: the 5D action for $\sigma$ is that of a free scalar as $z\to0$, the non-zero components of $F$ vanish as $z\to 0$, and $\sigma(z)$ satisfies the conditions in eq.~\eqref{eq:sigma-bc}.

\subsection{Other spacetimes}
\label{subsec:Model-others}
While the boundary condition for $\sigma$ in the IR is always true (when $\theta$ is related to a shrinking cycle), the action for $\sigma$ which almost has the form of a free scalar, requires that the spacetime be asymptotically of the form AdS$_5\times \mathbb{X}$, for some manifold $\mathbb{X}$. In this subsection we consider two other spacetimes to show this explicitly. While belaboring the point, it will be relevant for the later sections when we talk about the effect on the deconfinement to confinement phase transition. 

As a first example, let's consider AdS$_6$ as the starting point and make one of the spatial directions compact
\begin{align}
    \dd s^2 = \frac{\ell^2}{z^2}\left(-\dd t^2 + \dd \vec{x}^2 + \dd z^2 + \dd y^2\right)\:,\:\: y \cong y+ 2\cpi R\:.
\end{align}
We can restrict the $z$ direction to terminate at $z_\ir$, by hand, and impose the IR boundary condition $\sigma(z_\ir) = 0 $. Even though there is no ``disk'' in the spacetime, i.e. the compact dimension is not shrinking to zero size at $z = z_\ir$, we want to make this choice of boundary condition to make a fair comparison with the other cases. So far, there is effectively no difference from previous examples. However, the difference appears at the level of the action for $\sigma$: repeating the exercise of the previous subsection, one arrives at the action for $\sigma$ to be
\begin{align}
    S_5 = -\frac12 \int \dd^5 x \sqrt{g}\, f_\sigma^3 \, \frac{z}{\ell}\,(\partial\sigma)^2\:,\:\: f_\sigma^3 = \frac{1}{2\cpi e_6^2 R}\:.
\end{align}
The extra $z/\ell$ factor does not approach $1$ in the UV ($z\to0$), unlike before, and there is no limit in which we have a free scalar action. As we will see, this action will generate a contribution to the stabilizing potential that has to be effectively tuned away by choosing $\theta\to0$ if we do not want to destabilize the geometry.

One can wonder if the absence of an explicit ``disk'' is the culprit in the previous example. To address this, let us consider the example of AdS$_6$ soliton~\cite{Horowitz:1998ha} spacetime
\begin{align}
    \dd s^2 = \frac{\ell^2}{z^2}
    \left(
    \left(1-z^5/z_\ir^5\right)\dd y^2 + \dd \vec{x}^2 - \dd t^2 + \left(1-z^5/z_\ir^5\right)^{-1}\dd z^2 
    \right)\:,
\end{align}
which is obtained by a double analytic continuation of the extremal p-brane solution. The coordinate $y$ is periodic, $y \cong y+4\cpi z_\ir/5$, with the periodicity fixed by the requirement of smoothness of the geometry close to $z = z_\ir$. The ``disk'' here is in the $(z, y)$ subspace, with $z_\ir$ being the center of the disk. Doing a similar exercise as before, we run into an even worse problem: since the geometry on constant $y$ hypersurfaces is not AdS$_5$, starting with the 6D action for a gauge field, we do not get a 5D action of a scalar in AdS$_5$. If we ignore this, the action for $\sigma$ is of the form
\begin{align}
    S_5 = -\frac12 \int \dd^5 x \sqrt{g} f_\sigma^3 \frac{z}{\ell}\left(1-\frac{z^5}{z_\ir^5}\right)^{1/2}(\partial\sigma)^2\:,
\end{align}
where the metric is of the constant $y$ AdS$_6$ soliton geometry. Again the $z$ dependence is such that the action does not approach that of a free scalar in the UV ($z\to0$). In later sections, we will again see that the above action will generate a contribution to the stabilizing potential that has to be effectively tuned away by choosing $\theta\to0$, if we do not want to destabilize the geometry.  

\section{Radion Potential at Finite Vacuum Angle}
\label{sec:RadionPotential}

In this section we discuss the effect of a non-zero $\theta$ on the radion potential. We first present the standard results from the literature at $\theta = 0$, and then consider the effect of $\theta$ on each of them.

There is more than one choice for the radion potential, which is effectively fixed by the UV completion. Being in the 5D EFT, we have to make reasonable choices if we do not want to commit to a specific UV completion. These different choices come from different choices for $V(\Phi)$ and the boundary conditions for $\Phi$ on the UV/IR brane. In the 5D EFT, all these are equally valid choices, the requirement being that the radion potential should have a minimum and it should vanish at the minimum. These different choices for $\Phi$ dynamics correspond to choices of deformation of the CFT, which is an explicit breaking of CFT. The radion $\varphi$ acquiring a VEV is a spontaneous breaking of the CFT. The dynamics is therefore of a spontaneous breaking of CFT that is triggered by explicit breaking, and the different choices are parameterizing our ignorance in the ways this can happen.

\subsection{Radion potential at $\theta = 0$}
Without going into the details of the computation, we will consider three choices of parameters, which result in a radion potential that does the job it promises, but the underlying physics is different. The reason for these three choices will become clear soon. These three choices are coming from the choice of the bulk potential for $\Phi$, and the boundary conditions for $\Phi$ on the two branes. For all these cases, we will fix the boundary condition on the UV brane to be of Dirichlet type, requiring $\Phi(z = z_\uv) = v_\uv$, and stick to a purely quadratic potential for $\Phi$ in the bulk (see eq.~\eqref{eq:GW-bulk-potential}), but vary its sign. We will also add a detuning of the brane tensions when needed to have a minimum (i.e. the conditions in eq.~\eqref{eq:brane-tensions-and-bulk-cosmological-constant} are slightly violated). From this point onward $V(\varphi)$ denotes the full radion potential, including the additive constant $V_0$, chosen so that $V(\langle\varphi\rangle)=0$ at $\theta=0$. 

The details of the boundary condition on the 5D scalar $\Phi$, restrictions on the parameters, and the resulting radion potential for the three cases, which we call $\textbf{A, B, C}$ respectively, are:
\begin{align}
    &\textbf{A:}\qquad
    \Phi(z_\uv) = v_\uv\:,\:\Phi(z_\ir) = v_\ir\:,\: \epsilon > 0\:,\:v_\ir < v_\uv\:,
    \nonumber \\
    &\frac{\ell^4\,(V(\varphi)-V_0)}{24 \,M_5^3\ell^3} 
    = \frac{1+\epsilon/2}{6}\frac{v_\ir^2}{M_5^3}\,\varphi^4
    \left(
    \left(1 - \frac{v_\uv}{v_\ir} \varphi^\epsilon\right)^2 - \frac{\epsilon}{4+2\epsilon}  
    \right)
    \nonumber\\
    &\qquad\:\:\:\:\:\:\:\:
    = \frac{1+\epsilon/2}{6}\frac{v_\ir^2}{M_5^3}\,\varphi^4
    \left(
    \left(1 -  \frac{4+\epsilon+\sqrt{4\epsilon+\epsilon^2}}{4+2\epsilon}\left(\frac{\varphi}{\left<\varphi\right>}\right)^\epsilon\right)^2 - \frac{\epsilon}{4+2\epsilon}  
    \right)\:,
    \nonumber \\
    &\qquad\left<\varphi\right>
    \:= \left(\frac{v_\ir}{v_\uv}\right)^{1/\epsilon}\,\left(\frac{4+\epsilon+\sqrt{4\epsilon+\epsilon^2}}{4+2\epsilon}\right)^{1/\epsilon}
    \:. 
    \label{eq:radion-potential-type-A-theta-zero}\\[1em]
    &\textbf{B:}\qquad
    \Phi(z_\uv) = v_\uv\:,\:\Phi'(z_\ir) = \alpha_\ir\:,\: \epsilon > 0\:,\:\kappa < 0\:,\:\alpha_\ir v_\uv < 0\:,
    \nonumber \\
    &\frac{\ell^4\,(V(\varphi)-V_0)}{24 \,M_5^3\ell^3} 
    = \kappa\,\varphi^4
    -\frac{\alpha_\ir v_\uv}{48 M_5^3\ell^3}\,\varphi^{4+\epsilon}
    \nonumber\\
    &\qquad\:\:\:\:\:\:\:\:
    =  \kappa\,\varphi^4
    \left(
    1-\frac{1}{1+\epsilon/4}\left(\frac{\varphi}{\left<\varphi\right>}\right)^\epsilon
    \right)\:,
    \nonumber \\
    &\qquad\left<\varphi\right>
    \:= \left(\frac{48 M_5^3\ell^3\kappa}{(1+\epsilon/4)\alpha_\ir v_\uv}\right)^{1/\epsilon}
    \:. 
    \label{eq:radion-potential-type-B-theta-zero}\\[1em]
    &\textbf{C:}\qquad
    \Phi(z_\uv) = v_\uv\:,\:\Phi'(z_\ir) = \alpha_\ir\:,\: \epsilon < 0\:,\:\kappa > 0\:,\:\alpha_\ir v_\uv > 0\:,
    \nonumber \\
    &\frac{\ell^4\,(V(\varphi)-V_0)}{24 \,M_5^3\ell^3}
    = \kappa\,\varphi^4
    -\frac{\alpha_\ir v_\uv}{48 M_5^3\ell^3}\,\varphi^{4+\epsilon}
    \nonumber\\
    &\qquad\:\:\:\:\:\:\:\:
    = \kappa\,\varphi^4
    \left(
    1-\frac{1}{1+\epsilon/4}\left(\frac{\varphi}{\left<\varphi\right>}\right)^\epsilon
    \right)\:,
    \nonumber \\
    &\qquad\left<\varphi\right>
    \:= \left(\frac{48 M_5^3\ell^3\kappa}{(1+\epsilon/4)\alpha_\ir v_\uv}\right)^{1/\epsilon}
    \:. 
    \label{eq:radion-potential-type-C-theta-zero}
\end{align}
Here we have kept terms to leading order in $\varphi$, which is justified since we are interested in $\varphi\ll 1$ solutions (recall that $\varphi$ is dimensionless). For each case we have also written the potential in terms of $\left<\varphi\right>$ rather than in terms of the parameters, to make its behavior near the minimum $\varphi = \left<\varphi\right> $ transparent. The constant $V_0$ is chosen in each case to make the potential vanish at the minimum. Further, we have normalized the potential with an overall factor of $24 M_5^3\ell^3$, keeping the kinetic term for $\varphi$ in mind (see eq.~\eqref{eq:4d-action-gravity-and-radion}). The quantity $M_5^3\ell^3$ is related to the degrees of freedom of the dual CFT, and scales as $N^2$ for $SU(N)$ theory. The precise relation depends on the specific UV completion.\footnote{In type IIB AdS$_5\times S^5$ solution, $M_5^3\ell^3 = N^2/16\cpi^2$~\cite{Gubser:1999vj, Creminelli:2001th}. From thermodynamic considerations, $M_5^3\ell^3 = N^2/45\cpi^2$~\cite{Gursoy:2008za}. From requiring that the OPE of the glueball operator be correctly reproduced, $M_5^3\ell^3 = N^2/\cpi^2$~\cite{Csaki:2006ji}. These all differ by order one numbers. We will use $N^2/16\cpi^2$ in our estimates, but this ambiguity should be kept in mind.} Our normalization therefore pulls out an overall factor of $N^2$ out of the radion action: this is the ``glueball normalization''~\cite{Manohar:1998xv, Agashe:2019lhy}. The advantage of this normalization is that we can immediately read off that the quartic needs to be at most order 1, if we want to stay in the limit of small backreaction. For type $\mathbf{A}$ potential, this implies $v_\ir/M_5^{3/2} \lesssim 1$, and for types $\mathbf{B,C}$, it implies $\kappa \lesssim 1$.

The reason to choose these three forms of potential is because the interplay in various terms is different and once we switch on a $\theta$, the effect on the potential will be different. Let's first consider case $\textbf{A}$. The potential is a perfect square up to a term proportional to $\epsilon$. There are three extrema for this potential, and while there is a minimum at a nonzero $\varphi$, there is also one at $\varphi = 0$, which implies there is a barrier that separates the two. Working in the limit of small $\epsilon$, the barrier (the maximum of the potential) is located at $\varphi = e^{-1/\sqrt{\epsilon}}\left<\varphi\right> \ll \left<\varphi\right> $, and has a height of $e^{-4/\sqrt{\epsilon}}\,V_0 \ll V_0$. This barrier can be important when discussing the rate for the phase transition at very low temperatures. This barrier will also play a role in the stability of the minimum once a non-zero $\theta$ is allowed. 

\begin{figure}
    \centering
    \includegraphics[width=0.95\linewidth]{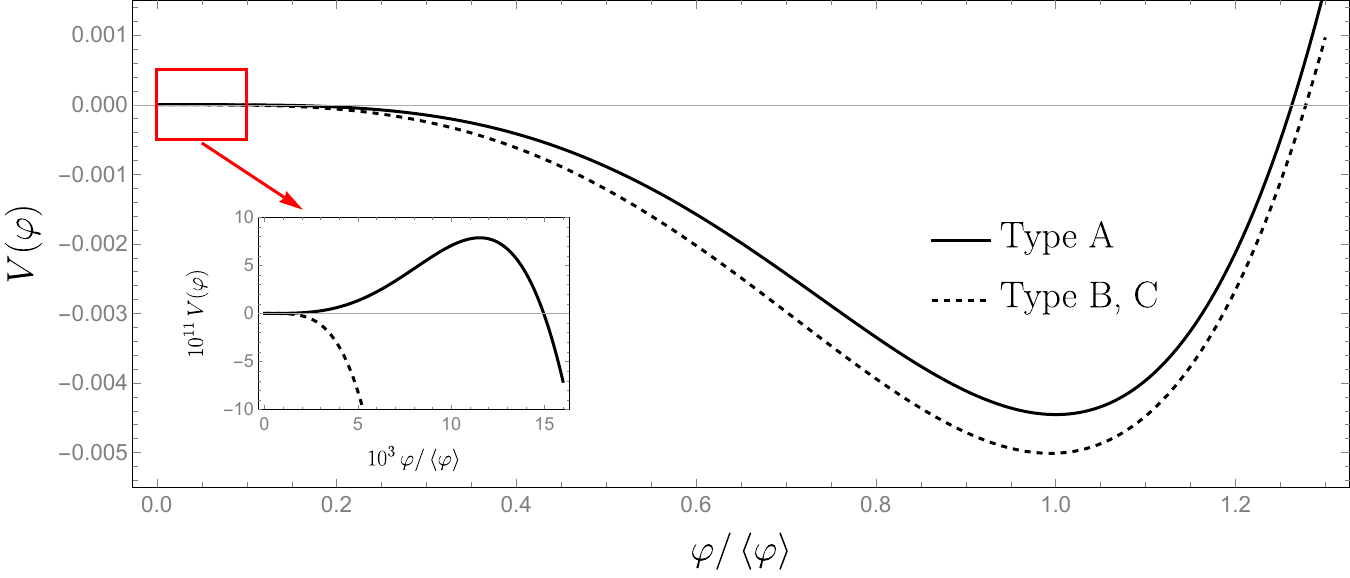}
    \caption{
    \small{Radion potential for the three cases discussed: type $\textbf{A}$ has a barrier at smaller $\varphi$, while type $\textbf{B, C}$ do not.}}
    \label{fig:RadionPotentials-type-ABC-theta-zero}
\end{figure}

Next consider case $\textbf{B}$. Here, there is only one minimum, and no barrier. Since $\epsilon > 0$, at small $\varphi$, $\varphi^4 \gg \varphi^{4+\epsilon}$. Choosing $\kappa < 0$ ensures the potential goes down as it starts from small $\varphi$, before it eventually rises up due to the $\varphi^{4+\epsilon}$ term with a positive coefficient, giving a minimum in the process. Case $\textbf{C}$ is similar: there is again only one minimum, no barrier. Since $\epsilon < 0$ now, for small $\varphi$, $\varphi^{4+\epsilon} \gg \varphi^4$, so the potential again goes down as it starts from zero (since the coefficient of $\varphi^{4+\epsilon}$ is negative), before it is overtaken by the $\varphi^4$ term with a positive coefficient, giving a minimum. The difference between case $\textbf{B}$ and $\textbf{C}$ is that in the first case the coefficient of the quartic is negative, while in the second case it is positive. We will see that this makes a difference when we discuss the effect of $\theta$. Figure~\ref{fig:RadionPotentials-type-ABC-theta-zero} shows the features of the three classes of radion potentials, for some choice of parameters. The bump for type $\textbf{A}$ is visible, and is absent for type $\mathbf{B, C}$, as explained. 

Unlike the IR boundary condition for $\sigma$, which was motivated above by regularity of the Wilson loop around a shrinking cycle, the IR data for the Goldberger--Wise field $\Phi$ are not fixed in the present bottom-up description. In a smooth top-down construction, the analog of these data would be determined by regularity and by the full coupled bulk solution, rather than by a freely chosen brane-localized potential. The three choices considered here should therefore be viewed as a parametrization of possible IR completions within the small-backreaction radion EFT. In particular, the loss of the confining minimum for some choices of parameters should be interpreted as a property of this effective description, and whether it actually occurs in a UV completion is an additional dynamical question, so it should be interpreted carefully.

\subsection{Contribution to the radion potential from $\theta$}
So far we have discussed the radion potential at $\theta = 0$. What about the contribution from a non-zero $\theta$? For this, we have to solve for the background profile for $\sigma$ in the geometry dual to the confined phase (eq.~\eqref{eq:metric-confined}). Recall that we are staying in the limit of small backreaction in the IR, which justifies using eq.~\eqref{eq:metric-confined}. After obtaining $\sigma(z)$, we plug it in the action and integrate over the $z$ coordinate to obtain the contribution to the potential. The equation of motion for $\sigma$ comes from minimizing the action in~\eqref{eq:5D-action-with-gravity-and-two-scalars}, and as explained earlier, we will choose $f_\sigma$ to be constant, and $V(\Phi,\sigma) = V(\Phi)$ to be independent of $\sigma$. The profile $\sigma(z)$ satisfies
\begin{align}
    \partial_z\left(\sqrt{g} g^{55} \partial_z\sigma \right) = 0\:,
    \label{eq:sigma-eom}
\end{align}
subject to the boundary conditions
\begin{align}
    \sigma(z_\uv) = \theta + 2 \cpi k\:,\:\: k \in \mathbb{Z}\:,\:\:\:\: \sigma(z_\ir) = 0\:.
    \label{eq:sigma-bc-explicit}
\end{align}
The profile for $\sigma$ is given as
\begin{align}
    \sigma(z) = (\theta + 2 \cpi k)\frac{z^4 - z_\ir^4}{z_\uv^4 - z_\ir^4}\:.
    \label{eq:sigma-profile-confined-phase}
\end{align}

Before proceeding to calculate the contribution to the radion potential, it is useful to pause and understand the solution we have obtained for $\sigma(z)$. Temporarily sending the UV brane to the boundary, i.e. $z_\uv \to 0$, we get
\begin{align}
    \sigma(z) = \left(\theta + 2\cpi k\right) - \left(\theta + 2\cpi k\right) (z/z_\ir)^4\:,
\end{align}
which using standard AdS/CFT dictionary, is interpreted as a source $\theta + 2\cpi k$ and a VEV $-\left(\theta + 2\cpi k\right)/z_\ir^4$. The presence of source is not a surprise, but the presence of a VEV, which is switched on because of the boundary condition $\sigma(z_\ir) = 0$ is the non-trivial aspect here. This boundary condition was motivated by how $\theta$ is generated in UV complete examples. As we will see, these aspects have a bearing on the discussions to follow. 

Plugging into the action and integrating over $z$, we get
\begin{align}
    S = -\frac12 f_\sigma^3 \int \dd^4 x \int_{z_\uv}^{z_\ir} \dd z \, \frac{\ell^3}{z^3}\left(\frac{4(\theta + 2 \cpi k) z^3}{z_\uv^4 - z_\ir^4}\right)^2  = 2 f_\sigma^3 \ell^3 \int \dd^4 x\frac{(\theta + 2 \cpi k)^2}{z_\uv^4 - z_\ir^4}\:.
\end{align}
To get the radion potential, we use the identification that $z_\uv/z_\ir = \varphi$ and promote it to a field $\varphi(x)$ which is the radion in eq.~\eqref{eq:4d-action-gravity-and-radion}. Naively, this would give the $\theta$ contribution to scale as $(\theta + 2\cpi k)^2$, where the integer $k$ is unspecified, and imply there are multiple branches. As we know that the dynamics should be invariant under $2\cpi$ shifts of $\theta$, we must further require a $\min k$ condition~\cite{Witten:1998uka}. This gives the $\theta$ contribution to the potential for $\varphi$ to be (taking $z_\uv = \ell$)
\begin{align}
    \ell^{4}V_\theta(\varphi) &= \min_k\:\:2 (f_\sigma \ell)^3\frac{(\theta + 2 \cpi k)^2 \varphi^4}{1-\varphi^4} \approx \min_k\:\: 2 (f_\sigma \ell)^3(\theta + 2 \cpi k)^2 \varphi^4\:,
    \nonumber \\
    &\equiv 24 (M_5\ell)^3\,\kappa_\theta \, \varphi^4\:\:,\:\:\:\: 
    \kappa_\theta = \min_k\, \frac{1}{12}\left(f_\sigma/M_5\right)^3(\theta + 2 \cpi k)^2\:,
    \label{eq:quartic-from-theta}
\end{align}
where the approximation in the first line is justified because we are interested in $\varphi\ll 1$. For notational brevity in what follows, in the second line we have defined $\kappa_\theta$ as a measure of the generated quartic, with appropriate factors of $(M_5\ell)^3$, keeping the normalization of eqs.~\eqref{eq:radion-potential-type-A-theta-zero},\eqref{eq:radion-potential-type-B-theta-zero},\eqref{eq:radion-potential-type-C-theta-zero} in mind. Since $M_5^3\ell^3 \sim N^2$, $\kappa_\theta$ scales as $1/N^2$.      

We see that we just get a quartic contribution to the radion potential from a non-zero $\theta$, to leading order in $\varphi$. Such a quartic can in fact also be generated by a simple detuning of the IR brane tension $\delta T_\ir$. Therefore, to leading order in $\varphi$, these two effects are degenerate. Conceptually there are differences: there is a multi-branch structure here with a minimization over $k$, the coefficient is proportional to square of an angle, is non-negative, and is suppressed at large $N$, features which are not present in the detuned IR tension case. Even more stark is the difference that $\theta$ is a UV quantity and $\delta T_\ir$ is an IR quantity. Practically, however, there is no difference.

We should also note that the $\min_k (\theta+2 \cpi k)^2$ factor in the generated potential immediately implies that the $k=0$ and $k = -1$ branches are degenerate at $\theta = \cpi$. The first derivative of the potential, proportional to $\min_k (\theta + 2\cpi k)$ is different at $\theta = \cpi$, for the $k = 0$ and $k = -1$ branches, leading to the familiar cusp in the potential at $\theta = \cpi$. This is the well-known signature of spontaneous CP breaking at $\theta = \cpi$. The order parameter for this breaking is proportional to $\partial_\theta V(\theta)$ and has a different value in the $k = 0$ and $k = -1$ branches. The bulk profile for $\sigma$ is also different in the two branches.

After the fact, a non-zero $\theta$ only generating a quartic contribution to the radion potential is not surprising: the dual operator $\tr F\widetilde{F}$ is marginal, and is therefore expected to generate a quartic on general scaling arguments. A crucial input comes from the boundary condition in the IR (see eq.~\eqref{eq:sigma-bc-explicit}), which ensures that a vacuum expectation value (VEV) for the operator is switched on, which changes with the energy scale (dual to the fact that $\sigma$ has a non-zero $z^4$ coefficient in eq.~\eqref{eq:sigma-profile-confined-phase}). This non-zero VEV, which comes from the IR boundary condition, and which further comes from $\theta$ being an angle, is the reason the effect is degenerate with a detuned IR brane tension: a very non-trivial interplay between various concepts. It was also crucial that there was no potential for $\sigma$, for arriving at this conclusion. The exercise has however established that the coefficient of the quartic is positive, $\kappa_\theta > 0$ and it scales as $\theta^2$, a fact that will be relevant in what follows. 

A positive $\kappa_\theta$ means that in isolation, the effect of a non-zero $\theta$ is to drive the radion field $\varphi$ to zero. There is a physical explanation for this. Recall that we generated $\theta$, at least in the examples we considered, by switching on a flux in the higher-dimensional realization. Fluxes in a compact space generically want to make the space bigger, driving towards decompactification, which is the $\varphi\to0$ limit here (recall that $\varphi = z_\uv/z_\ir$ and a larger $z_\ir$ means a larger fifth dimension). 

One can also derive the contribution to the radion potential directly, by looking at the on-shell action of the gauge fields that generate the 5D field $\sigma$. For the 6D example in sec.~\ref{subsec:Model-6D}, evaluating the action on the background solution for $A_6$ (eq.~\eqref{eq:model-6D-action-and-profiles}) we get
\begin{align}
    S_6 &\supset -\frac{1}{2e_6^2} \int \dd^6 x \sqrt{g_6} (\partial_A A_6)(\partial_C A_6) g_6^{AC} g_6^{66} 
    \nonumber \\
    &= -\frac{1}{2 e_6^2}
    \int \dd^4 x
    \int_0^{2\cpi\ell} \dd y 
    \int_{z_\uv}^{z_\ir} \dd z 
    \frac{\ell^3}{z^3}\left(1-\frac{z}{z_\ir}\right)^{-1}\left(\frac{\dd A_6}{\dd z}\right)^2
    \nonumber \\
    &=-\frac{2\cpi\ell}{2 e_6^2}\left(\frac{\theta + 2 \cpi k}{2\cpi \ell}\right)^2\int \dd^4 x \int_{z_\uv}^{z_\ir} \dd z \frac{\ell^3}{z^3}\left(1-\frac{z}{z_\ir}\right)
    \left(\frac{20z^3}{z_\ir^4}\right)^2 \:.
\end{align}
Again identifying $\varphi = z_\uv/z_\ir$, taking $z_\uv = \ell$, and keeping to leading order in $\varphi$, we get
\begin{align}
    V_\theta(\varphi) = \min_k \, \frac{5}{\cpi}\frac{1}{e_6^2\ell^2}\left(\theta+2\cpi k\right)^2\varphi^4 
    = \min_k \, 10\, (f_\sigma^3/\ell)(\theta+2\cpi k)^2\varphi^4\:,
\end{align}
and we again get a pure quartic. The overall factor is different than in eq.~\eqref{eq:quartic-from-theta}, and is theory specific. Even though the action for $\sigma$ was not that of a free scalar in this case (see eq.~\eqref{eq:model-6D-action-and-profiles}), it does not change the scaling of the leading order contribution to the radion potential.

For the two spacetimes considered in sec.~\ref{subsec:Model-others}, the same exercise leads to a $\varphi^3$ (for 6D AdS) and a $\varphi^2$ (for 6D AdS Soliton) potential. Since we are looking for $\varphi \ll 1$ solutions and the $\theta = 0$ potential balances terms that are approximately quartic ($\varphi^4$ vs $\varphi^{4+\epsilon}$ etc), a term like $\varphi^2$ or $\varphi^3$, unless aggressively fine-tuned, will generically destabilize the geometry. This is closely tied to the fact that for these spacetimes, we did not find the low energy 5D action for $\sigma$ to approach that of a free scalar, even asymptotically in the UV. In what follows we will not consider these spacetimes. 

Before proceeding to discuss the full radion potential and other quantities relevant to the phase transition, we would like to make a brief remark about the special situation at $\theta = \cpi$. As derived earlier, the potential, being proportional to $\text{min}_k \left(\theta + 2 \cpi k\right)^2$ is degenerate at $\theta = \cpi$, for the $k = 0$ and $k = -1$ branches. The two branches have opposite values of the CP-odd order parameter
$\partial_\theta V \, \propto \, \theta+2\cpi k$, and are exchanged by CP. Thus one expects a stable planar domain wall interpolating between the two CP-conjugate vacua. In the five-dimensional description, the bulk part of such a wall would be described by a configuration $\sigma(x,z)$, where $x$ is the coordinate transverse to the wall. Far from the wall it approaches the two homogeneous profiles with UV values $+\cpi$ and $-\cpi$, while both sides obey the same IR condition of vanishing $\sigma$. The physical UV source is not varying: $e^{i\sigma_\uv}=-1$ everywhere. The apparent jump is only a jump in a lift of the compact field, or equivalently a change of branch. At the level of the classical scalar equation, choosing a representative lift of the UV data gives a bulk flux profile whose radial conjugate momentum, dual to the CP-odd operator, changes sign across the wall.

A complete description of the wall requires additional localized data beyond the minimal scalar profile. This expectation comes from pure $SU(N)$ Yang-Mills theory, where the mixed 't Hooft anomaly between CP/time-reversal and the one-form center symmetry $\mathbb{Z}_N^{(1)}$ at $\theta = \cpi$ implies that the wall supports an $SU(N)_1$ Chern-Simons TQFT~\cite{Gaiotto:2017yup}. In the simplified five-dimensional model this sector would have to be supplied by the appropriate branch-changing core or localized wall degrees of freedom. The wall with a localized TQFT has interesting properties regarding the behavior of Wilson lines and confining strings, and it would be instructive to study some of them in the simplified model presented here. We leave these aspects for future work.

\subsection{Radion potential at $\theta\neq0$}

Let us now consider the effect of an additional positive quartic on the radion potentials of types $\mathbf{{A,B,C}}$. 
We denote the minimum of the radion potential at $\theta = 0$ by $\left<\varphi\right>$, and at a non-zero $\theta$ by $\left<\varphi\right>_\theta$.

\begin{figure}
    \centering
    \includegraphics[width=0.95\linewidth]{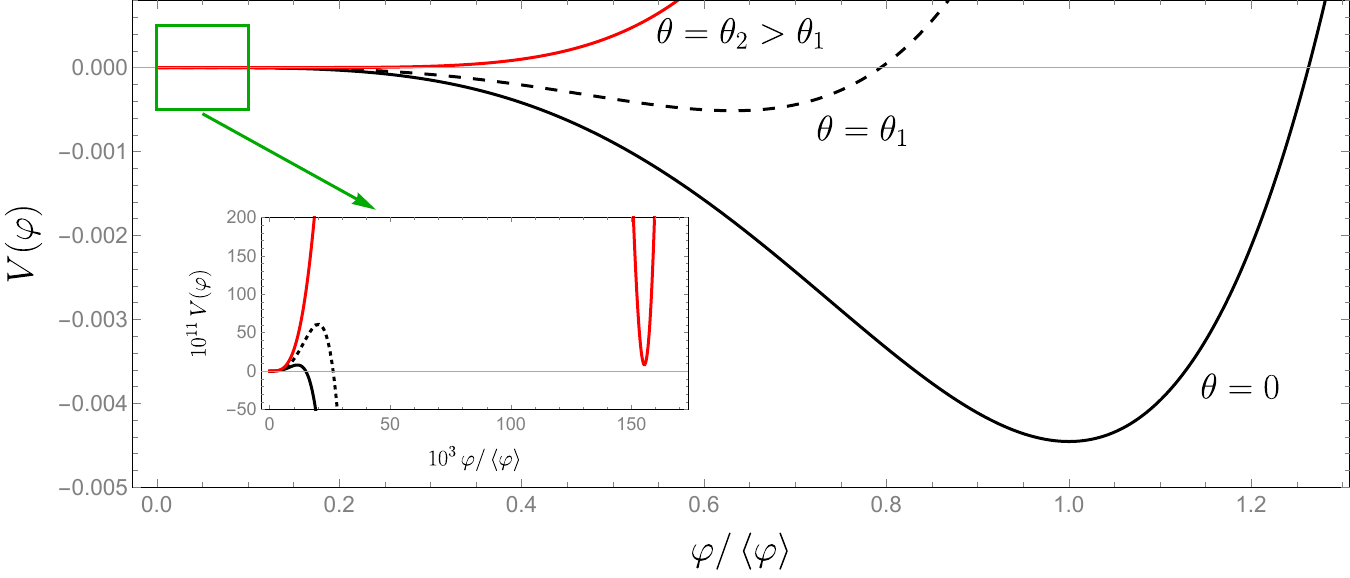}
    \caption{
    \small{Effect of $\theta$ on the radion potential of type $\textbf{A}$.}
    }
    \label{fig:RadionPotentials-type-A-theta-non-zero}
\end{figure}

For type $\mathbf{A}$, which has a barrier between $\varphi = 0$ and $\left<\varphi\right>$, the effect of a positive quartic is to move the minimum towards smaller values, $\left<\varphi\right>_\theta < \left<\varphi\right>$, while increasing the barrier. This is shown in fig.~\ref{fig:RadionPotentials-type-A-theta-non-zero}. At a critical value of $\theta$, the two extrema merge and there is no minimum any more! Therefore to have a minimum, $\kappa_\theta$ must satisfy
\begin{align}
    \kappa_\theta < \kappa_\theta^* =  \frac{\epsilon \, v_\ir^2}{24 M_5^3} \left(1+\frac{\epsilon}{4}\right)\:.\qquad \text{(type $\mathbf{A}$)}
    \label{eq:kappa-restriction-typeA}
\end{align}
Using eq.~\eqref{eq:quartic-from-theta} and $(M_5\ell)^3 = N^2/16\cpi^2$, stability translates to a condition on $N$:
\begin{align}
    N >  8\sqrt{2}\,\cpi^2(f_\sigma\ell)^{3/2}\,\left|\frac{\theta}{2\cpi}+k\right|\,
    \frac{M_5^{3/2}/v_\ir}{\epsilon^{1/2}(1+\epsilon/4)^{1/2}}\:.
\end{align}
For $f_\sigma\ell = 1, \epsilon = 10^{-1}, v_\ir/M_5^{3/2} = 1, \theta/2\cpi = 0.1$, the above gives $N > 35$ (for the principal branch near $\theta = 0$ where $k = 0$), which is a very strong requirement! Alternatively, one can read this condition to restrict $\theta$, given $N$. E.g. for $N = 10, \epsilon = 10^{-1}, f_\sigma\ell = 1, v_\ir/M_5^{3/2} = 1$, we need $\left|\theta/2\cpi\right| < 0.03$ (for the principal branch near $\theta = 0$ where $k = 0$). For different UV completions, these restrictions are expected to be affected by order 1 numbers.

For radion potential of type $\textbf{B,C}$ which only have one minimum and no barrier, the effect of a non-zero $\theta$ is again to reduce $\left<\varphi\right>$, $\left<\varphi\right>_\theta < \left<\varphi\right>$. This is straightforward to understand: the form of the potential stays the same when $\theta$ is non-zero, if one simply replaces $\kappa$ by $\kappa + \kappa_\theta$. The minimum of the potential at $\theta = 0$ is at 
\begin{align}
    \left<\varphi\right> = \left(\frac{48 M_5^3\ell^3 \kappa}{(1+\epsilon/4)\alpha_\ir v_\uv}\right)^{1/\epsilon}    
\end{align}
For both type $\mathbf{B,C}$, the ratio $\kappa/(\alpha_\ir v_\uv) > 0$. Depending on the sign of $\epsilon$, we need this ratio to be larger or smaller than 1, to generate $\left<\varphi\right> \ll 1$. 

For type $\mathbf{B}$ with $\epsilon > 0$, we need $\kappa/(\alpha_\ir v_\uv) \lesssim 1$. Since $\kappa < 0$ and $\kappa_\theta > 0$, the overall $\kappa$ reduces (in magnitude), which makes the ratio $\kappa/(\alpha_\ir v_\uv)$ smaller and hence $\left<\varphi\right>$ smaller, resulting in $\left<\varphi\right>_\theta < \left<\varphi\right>$. To have a minimum, we again need $\kappa_\theta$ to satisfy
\begin{align}
    \kappa_\theta < -\kappa\:.\qquad \text{(type $\mathbf{B}$)}
    \label{eq:kappa-restriction-typeB}
\end{align}
Using eq.~\eqref{eq:quartic-from-theta} and $(M_5\ell)^3 = N^2/16\cpi^2$, the corresponding condition on $N$ is:
\begin{align}
    N > \frac{4}{\sqrt{3}}\cpi^2\,\frac{(f_\sigma\ell)^{3/2}}{\left|\kappa\right|^{1/2}}\,\left|\frac{\theta}{2\cpi} + k \right|\:.
\end{align}
For $f_\sigma\ell = 1, \kappa = -0.1, \theta/2\cpi = 0.1$, we need $N>7$ (for the principal branch near $\theta = 0$ where $k = 0$). Alternatively, one can again read this condition to restrict $\theta$, given $N$. E.g. for $N = 10, \kappa = -0.1, f_\sigma \ell = 1$, we need $\left|\theta/2\cpi\right| < 0.14$ (for the principal branch near $\theta = 0$ where $k = 0$). For different UV completions, these restrictions are expected to be affected by order 1 numbers. 

For type $\mathbf{C}$ with $\epsilon < 0$, we need $\kappa/(\alpha_\ir v_\uv) \gtrsim 1$. Since $\kappa > 0$ and $\kappa_\theta > 0$, the overall $\kappa$ increases (in magnitude), which makes the ratio $\kappa/(\alpha_\ir v_\uv)$ larger and hence $\left<\varphi\right>$ again smaller (since this time it is raised to a large negative power, $\epsilon$ being negative). Therefore one again gets $\left<\varphi\right>_\theta < \left<\varphi\right>$. Unlike type $\textbf{B}$, there is no restriction on $\kappa_\theta$ this time.  

\begin{figure}
    \centering
    \includegraphics[width=0.9\linewidth]{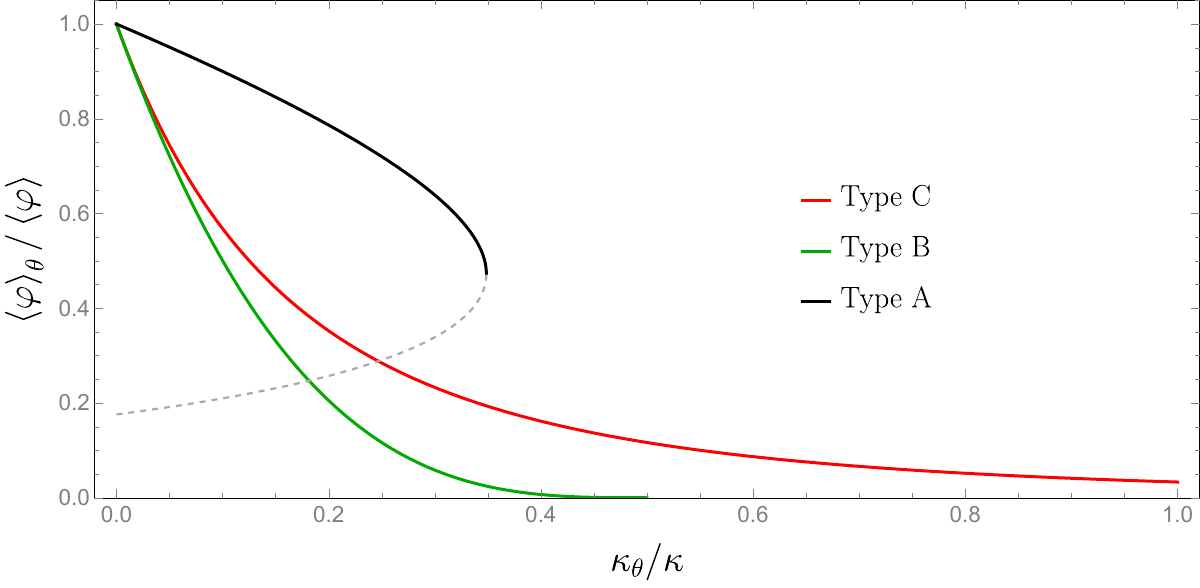}
    \caption{
    \small{Location of minimum as a function of $\kappa_\theta$ for type $\mathbf{A, B, C}$ potentials in black, red, green respectively. Dashed gray line shows the maximum that merges with the minimum for type $\mathbf{A}$ type potential.}
    }
    \label{fig:MovingMinimum-type-ABC}
\end{figure}

We thus see that in all cases, $\left<\varphi\right>$ reduces due to a non-zero $\theta$. However there is one crucial difference between these cases---for type $\mathbf{B}$, the overall $\kappa$ can become zero, which makes $\left<\varphi\right>_\theta \to 0$. Like type $\mathbf{A}$, the minimum disappears at a critical value  $\kappa_\theta$. However, unlike type $\mathbf{A}$, the minimum approaches zero before it disappears. There is no such effect for type $\mathbf{C}$. In summary, one can lose the minimum for type $\mathbf{A}$ and type $\mathbf{B}$ potentials if $\kappa_\theta$ is large enough, but not for type $\mathbf{C}$. When the minimum is lost, $\left<\varphi\right>_\theta$ is arbitrarily small for $\mathbf{B}$, but is finite for type $\mathbf{A}$. Figure~\ref{fig:MovingMinimum-type-ABC} shows the ratio $\left<\varphi\right>_\theta/\left<\varphi\right>$ as a function of $\kappa_\theta$ for the three cases. We remind the reader that these results are derived in a bottom up model and $\mathbf{A, B, C}$ are parameterizing our ignorance. It would be useful to see how these conclusions are changed if the brane localized parameters $\alpha_\ir, v_\uv, \kappa$ get related in a more UV complete model. 

The VEV $\left<\varphi\right>$ is related to the confinement scale $\Lambda_c$ in the dual theory. One can take the mass of the lightest glueball $m \, \propto \, \Lambda_c$, or the string tension $T_s \, \propto \, \Lambda_c^2$ to be a placeholder for $\Lambda_c$. Both $m$ and $T_s$ are measured on the lattice as a function of $\theta$~\cite{Bonanno:2024ggk} and are given as
\begin{align}
    &\frac{m(\theta)}{m(0)} = 1-\frac{m_2}{N^2}\theta^2 + \mathcal{O}(\theta^4)\:,\:\:
    \frac{T_s(\theta)}{T_s(0)} = 1-\frac{s_2}{N^2}\theta^2 + \mathcal{O}(\theta^4)\:,
    \nonumber
    \\
    & s_2 = 0.23(1)\:,\:\: m_2 = 0.075(20)\:\:.
\end{align}
The crucial aspect is that both $m$ and $T_s$ decrease as a function of $\theta$ and the effect vanishes in the large $N$ limit. This in fact is the same behavior seen in $\left<\varphi\right>_\theta/\left<\varphi\right>$ for each of the three choices of potentials $\mathbf{A, B, C}$! While it is tempting to make this comparison more quantitative, we will not do it yet, since $m \sim \left<\varphi\right>, T_s \sim \left<\varphi\right>^2$ are only qualitative conditions at best, and since $s_2, m_2$ are not equal, one has to do a precise calculation of $m, T_s$ in the 5D model to make the comparison meaningful. Nonetheless, it is encouraging to see the qualitative agreement. A precise calculation of $m, T_s$ at non-zero $\theta$ and a comparison with the lattice result will be left for future work. 

\section{Critical Temperature at Finite Vacuum Angle}
\label{sec:CriticalTemperature}
Having seen the effect of $\theta$ on the radion potential, it is straightforward to calculate the effect on the critical temperature. We first need to calculate the contribution to the free energy of the deconfined and confined phases from the $\sigma$ action. One way to calculate the free energy is to look for solutions to the Euclidean action $S_E$ with periodic time $t_E\cong t_E+ 1/T$, and evaluate the action on-shell. The free energy $f$ (per unit 3-volume) is given by $f = T S_E$.  For the confined phase, since the solution is time independent and corresponds to $\varphi = \left<\varphi\right>_\theta$, the free energy is simply given by the value of the radion potential at the minimum (the factor of $T$ cancels with the $1/T$ from integral over $t_E$):
\begin{align}
    f_c = V(\left<\varphi\right>_\theta) \:,
\end{align}
and is non-zero in general, since $V_0$ in eqs.~\eqref{eq:radion-potential-type-A-theta-zero},\eqref{eq:radion-potential-type-B-theta-zero},\eqref{eq:radion-potential-type-C-theta-zero} were chosen to ensure $V(\left<\varphi\right>) = 0$, and $V(\left<\varphi\right>_\theta)\neq V(\left<\varphi\right>)$. Note that the same constant $V_0$ also appears on the deconfined side~\cite{Agashe:2019lhy}. 

To obtain the contribution to the free energy on the deconfined side, we need to find the solution for $\sigma(z)$ in the black brane background, subject to the boundary conditions $\sigma(z_\uv) = \theta + 2\cpi k$ and regularity at the horizon $z = z_h$ (i.e. the field value and derivatives are finite there). The equation of motion for $\sigma$ from eq.~\eqref{eq:sigma-eom}, in the black brane background, is given as
\begin{align}
    \frac{\dd}{\dd z} \, \left(\frac{\ell^3}{z^3}\left(1-\frac{z^4}{z_h^4}\right)\,\frac{\dd \sigma}{\dd z}\right) = 0\:,
\end{align}
whose solution is
\begin{align}
    \sigma(z) = A + B \log (1-\frac{z^4}{z_h^4})\:,
\end{align}
for constant $A,B$. Since the log term is singular at $z = z_h$, regularity requires $B = 0$. Using the UV boundary condition, we get
\begin{align}
    \sigma(z) = \theta + 2 \cpi k\:,\:\: k \in \mathbb{Z}\:,
    \label{eq:sigma-profile-deconfined-phase}
\end{align}
which is independent of $z$. The on-shell action is then zero (up to the additive constant $V_0$), since it only involves derivatives of $\sigma$. 

In conclusion, the change in the free energy from a non-zero $\theta$ appears only in the confined phase. At finite $\theta$, we therefore have
\begin{align}
    f_d - f_c &= \left(-2\cpi^4 (M_5\ell)^3 T^4 + V_0\right) - \left(V(\left<\varphi\right>_\theta)\right) 
    \nonumber \\
    &= -2\cpi^4 (M_5\ell)^3 T^4 
    +V_0 - V(\left<\varphi\right>_\theta)
\end{align}
At the critical temperature, the free energies of the two phases are equal, which gives
\begin{align}
    T_c(\theta) = \left(\frac{V_0-V(\left<\varphi\right>_\theta)}{2\cpi^4 (M_5\ell)^3}\right)^{1/4}\:.
    \label{eq:tc-theta-in-terms-of-potential}
\end{align}
Using eq.~\eqref{eq:critical-temperature-no-theta}, we get
\begin{align}
    T_c(\theta) = T_c(0)\,\left( 1- \frac{V(\left<\varphi\right>_\theta)}{V_0}\right)^{1/4}
    \label{eq:Tc-at-theta-general}
\end{align}
For potentials of type $\mathbf{B,C}$, we get (taking $k = 0$ branch)
\begin{align}
    \frac{T_c(\theta)}{T_c(0)} &= \frac{\left<\varphi\right>_\theta}{\left<\varphi\right>\,}
    \left( 1+\frac{\kappa_\theta}{\kappa}\right)^{1/4}
    = 
    \left( 1+\frac{\kappa_\theta}{\kappa}\right)^{1/4+1/\epsilon}
    \nonumber \\
    &= \left(1+\frac{1}{12}\frac{f_\sigma^3}{M_5^3}\frac{\theta^2}{\kappa}\right)^{1/4+1/\epsilon}
     \nonumber \\
     &= 
     \left(1+\frac{16\cpi^4}{3}\frac{f_\sigma^3\ell^3}{\kappa\, N^2}\left(\frac{\theta}{2\cpi}\right)^2\right)^{1/4+1/\epsilon}
    \:,
\end{align}
where we have used eq.~\eqref{eq:quartic-from-theta} in the second line to write $\kappa_\theta$ in terms of $f_\sigma, M_5, \ell$, and used $M_5^3\ell^3 = N^2/16\cpi^2$ in the third line to show the $N$ dependence explicitly. Note that for type $\mathbf{B}$, when $\kappa_\theta + \kappa \to 0$, the critical temperature $T_c \to 0$. For type $\mathbf{C}$, $\kappa_\theta + \kappa$ is always positive, so $T_c$ is always non-zero.

We can expand the expression for $T_c(\theta)/T_c(0)$ to get
\begin{align}
    \frac{T_c(\theta)}{T_c(0)} &= 
    \left( 1+\frac{\kappa_\theta}{\kappa}\right)^{1/4+1/\epsilon}
    = \left[ 1+\left(\frac{1}{\epsilon}+\frac14\right)\frac{\kappa_\theta}{\kappa} + \frac12\left(\frac{1}{\epsilon}+\frac14\right)\left(\frac{1}{\epsilon}-\frac34\right)\frac{\kappa_\theta^2}{\kappa^2} + \mathcal{O}(\kappa_\theta^3)\right]
    \nonumber \\
    &= \left(1-\frac{16\cpi^4}{3}\frac{f_\sigma^3\ell^3}{\left|\epsilon\,\kappa\right|\, N^2} \left(\frac{\theta}{2\cpi}\right)^2 
    + 
    \frac{128\cpi^8}{9}\frac{f_\sigma^6\ell^6}{\epsilon^2\,\kappa^2\, N^4} \left(\frac{\theta}{2\cpi}\right)^4 + \mathcal{O}(\theta^6)
    \right)
    \:,
\end{align}
where we have used $\left|\epsilon\right| \ll 1$ and $\epsilon\,\kappa < 0$ in the second line. As before, we have used eq.~\eqref{eq:quartic-from-theta} to write $\kappa_\theta$ in terms of $f_\sigma, M_5, \ell$ and used $M_5^3\ell^3 = N^2/16\cpi^2$. 

We see that for small $\theta$, the correction scales with $\theta^2/N^2$ and decreases $T_c$. This is parametrically consistent with the results seen on the lattice~\cite{DElia:2012pvq, DElia:2013uaf, Bonati:2016tvi, Bonanno:2023hhp}
\begin{align}
    T_c(\theta)/T_c(0)  =  1- R_\theta\, \theta^2/N^2\:,\:\: R_\theta = 0.159(4)\:.
\end{align}
For a precise numerical comparison, we need to specify the model parameters $\kappa, \epsilon, f_\sigma$, and also the relation between $M_5$ and $N$.
In the next section where we calculate the topological susceptibility, we will be able to give a more model-independent comparison with the lattice data. At the moment, the more important thing is the sign and the scaling with $\theta$ and $N$. The full expression also shows the scaling of the higher order terms, and in particular that the coefficient of $\theta^4$ is positive, unlike the coefficient of $\theta^2$. This is also seen in UV complete models~\cite{Bigazzi:2015bna}.

For potential of type $\mathbf{A}$, there are a few differences. First is that $\epsilon$ is always positive. Second, the coefficient of the quartic term at $\theta = 0$ is $v_\ir^2/(6M_5^3)$ (see eq.~\eqref{eq:radion-potential-type-A-theta-zero}), and is not a free parameter. Finally, we have to be careful in restricting $\kappa_\theta$ to be in the range when the minimum does not get destabilized. 
Staying away from the boundary of this condition, we can use $\left<\varphi\right>_\theta = \left<\varphi\right>$ since it does not change the result for $T_c(\theta)/T_c(0)$ to leading order.\footnote{Close to the boundary of the stability condition in eq.~\eqref{eq:kappa-restriction-typeA}, the minimum has higher free energy than the one at $\varphi = 0$.} We get (again taking $k=0$ branch)
\begin{align}
    \frac{T_c(\theta)}{T_c(0)} &= \frac{\left<\varphi\right>_\theta}{\left<\varphi\right>\,}\,
    \left(
    1-\frac{24 M_5^3\,\kappa_\theta}{v_\ir^2\,\epsilon}
    \right)^{1/4}
    =\:\: 
    \left[
    1-\frac{16\cpi^4}{3}\frac{f_\sigma^3\ell^3}{\epsilon\, N^2\,(v_\ir^2/6M_5^3)}\left(\frac{\theta}{2\cpi}\right)^2
    +\mathcal{O}(\theta^4)\right]
    \:.
\end{align}
where we have expanded to linear order, used $\left<\varphi\right>_\theta = \left<\varphi\right>$, and written the expression in a form that allows identifying the ``$\kappa$'' of type $\mathbf{A}$ potential, $v_\ir^2/6M_5^3$ (see eq.~\eqref{eq:radion-potential-type-A-theta-zero}), making the final result in the second line be of the same form as for type $\mathbf{B, C}$.

In summary, the simple 5D model that was constructed in a bottom up way is able to capture the parametric dependence of the critical temperature for the deconfinement to confinement phase transition at non-zero $\theta$, as seen on the lattice~\cite{DElia:2012pvq, DElia:2013uaf, Bonati:2016tvi, Bonanno:2023hhp}, and in other UV models~\cite{Bigazzi:2015bna, Bartolini:2016dbk}. Having this simple model now allows us to go further than lattice and compute the rate for the phase transition, at least in the range of parameters where the model is sensible.
  
\section{Topological Susceptibility across the Critical Temperature}
\label{sec:TopologicalSusceptibility}
The effect of a non-zero background value of $\theta$, which sources the operator $F\widetilde{F}$, can be measured by the two-point function of the operator $F\widetilde{F}$. The IR (zero-momentum) limit of this two-point function is proportional to the topological susceptibility $\chi$. It is straightforward to calculate $\chi$ if one knows the vacuum energy as a function of $\theta$ (in a given phase), using $\chi = \partial_\theta^2V(\theta)$. In our calculations so far, we have already calculated $V(\theta)$ for the deconfined and the confined phases, using the on-shell action, so the calculation of $\chi$ easily follows. 

For the confined phase, from the expression for $V_\theta$ in eq.~\eqref{eq:quartic-from-theta} one can immediately read off the topological susceptibility to be (keeping to $k=0$ branch)
\begin{align}
    \chi_c = \partial_\theta^2 V_\theta = 4(f_\sigma\ell)^3(\varphi/\ell)^4\:,
\end{align}
which is dependent on the radion $\varphi$. We get $\chi_c \:\: \propto \:\: N^0$, as expected from large $N$ counting. For a stabilized geometry, when $\varphi$ has a VEV, $\chi_c \:\: \propto \:\:  (\left<\varphi\right>/\ell)^4 \sim \Lambda_c^4$, where $\Lambda_c$ is the confinement scale. It is useful to point out that in many early Universe scenarios, $\varphi$ can have a time-dependent background solution~\cite{Mishra:2025ofh}, which $\chi_c$ will inherit. Since many aspects of the spectrum and dynamics depend on $\chi_c$, there can be interesting consequences of such a time-dependence.

For the deconfined phase, we saw that regularity at the horizon forced $\sigma$ to have a constant profile and the on-shell action was independent of $\theta$ (see discussions near eq.~\eqref{eq:sigma-profile-deconfined-phase}). As a consequence, the topological susceptibility in the deconfined phase is zero:
\begin{align}
    \chi_d = 0\:.
\end{align}
We should note that there can be a bulk potential for $\sigma$ if one includes  instanton effects. This effect would be suppressed by $e^{-N}$, and so would be $\chi_d$. We have of course ignored these effects.

We therefore see that $\chi$ has a sharp drop at the critical temperature $T_c$, being approximately constant for $T< T_c$ and zero for $T>T_c$. This behavior is seen on the lattice~\cite{DelDebbio:2004vxo, Vicari:2008jw}.\footnote{As mentioned, instanton effects can generate a bulk potential for $\sigma$ which would be suppressed by $e^{-N}$. The profile of $\sigma$ would not be flat anymore in the deconfined phase, and $\chi_d$ would not be zero anymore, but rather suppressed by $e^{-N}$. This explains why the drop in $\chi$ as $T\to T_c$ from below, as seen on the lattice, gets sharper as $N$ increases.} A priori, this is a peculiar behavior: why should the $\theta$ dependence care about the critical temperature, or effectively about confinement? In the holographic setup we have considered, the answer is very simple: it is because $\theta$ came from winding on a compact spatial direction, the same one which shrank to zero size in the IR, capping off the geometry and signaling confinement.

Having calculated the topological susceptibility for the confined phase, we can now go back to the behavior of critical temperature $T_c$ as a function of $\theta$, and present our result in a way that circumvents specifying model parameters like $\epsilon, \kappa$ etc, making the result more model-independent. Starting with eq.~\eqref{eq:Tc-at-theta-general} and using the definition of topological susceptibility $V(\left<\varphi\right>_\theta) = \chi_c(0) \theta^2/2 + \mathcal{O}(\theta^4)$, we can write
\begin{align}
    \frac{T_c(\theta)}{T_c(0)} &= \left( 1- \frac{V(\left<\varphi\right>_\theta)}{V_0}\right)^{1/4} \approx \left( 1- \frac{\chi_c(0) \theta^2}{2V_0}\right)^{1/4} \approx 1-\frac{1}{8}\frac{\chi_c(0)}{V_0}\theta^2
    \nonumber
    \\
    &= 1-\frac{1}{16\cpi^4 (M_5\ell)^3}\frac{\chi_c(0)}{T_c(0)^4}\theta^2\:,
\end{align}
where in the first line we have kept to first order in $\theta^2$ and in the second line we have used the relation $V_0 = 2 \cpi^4 (M_5 \ell)^3 T_c(0)^4$ (e.g. obtained by setting $\theta = 0, V(\left<\varphi\right>_\theta) = 0$ in eq.~\eqref{eq:tc-theta-in-terms-of-potential}). Comparing with the lattice parametrization $T_c(\theta)/T_c(0) = 1-R_\theta \theta^2/N^2$ we get
\begin{align}
    R_\theta = \frac{1}{16\cpi^4}\frac{N^2}{(M_5\ell)^3}\frac{\chi_c(0)}{T_c(0)^4}\:.
\end{align}
As promised, we have been able to write the result without needing to specify $\epsilon, \kappa, f_\sigma$ etc. The only ambiguity comes from the relation between $M_5$ and $N$. In fact we can use the lattice result to fix this relation. Lattice results quote $T_c(0)/\sqrt{T_s} = 0.5949(17)$~\cite{Lucini:2012wq} and $\chi_c(0)/T_s^2 = 0.02088(39)$~\cite{Bonanno:2025eeb}, where $T_s$ is the zero-temperature string tension. From this we get $\chi_c(0)/T_c(0)^4 \approx 0.17$. Using $R_\theta = 0.159(4)$ from the lattice we get
\begin{align}
    (M_5 \ell)^3 = 0.11 \, \frac{N^2}{16 \cpi^2} = 0.31 \, \frac{N^2}{45 \cpi^2}\:,
\end{align}
where we have shown the pre-factor for the two popular choices of relation between $M_5$ and $N$. 

Rather than leaning too much into the numerical factors, the robust conclusion that can be drawn at this point is that the simplified 5D model captures the parametric structure, sign, and large-$N$ scaling of the lattice result. This allows using the model to compute quantities that are difficult to calculate on the lattice. In the next section, we use this model to calculate the transition rate at non-zero vacuum angle.  

\section{Transition Rate at Finite Vacuum Angle}
\label{sec:TransitionRate}

With an explicit 5D model that incorporates the $\theta$ term, we can make quantitative statements about the dynamics of the deconfined to confined phase transition.  
The general calculation for the rate of transition from the deconfined to the confined phase is quite subtle, and requires making additional assumptions about the dynamics. In the gravitational dual, one has to identify a gravitational instanton that takes the black brane spacetime to the spacetime with an IR brane. This instanton also has to change the topology of the spacetime, which is easiest seen in the Euclidean picture where the black brane geometry has a disk factor and the IR brane geometry has an annulus factor. It is currently an open problem to construct the instanton in the full 5D description. The currently used approach in the literature is to use the 4D KK reduced EFT of the radion, which is only justified if the radion is lighter than the KK scale. Even in this limit, since the instanton in general involves field values of the radion where the KK scale gets lowered (compared to the temperature), it is not entirely in full theoretical control and one has to make additional assumptions. 

To make progress, primarily two approaches exist. In the first approach, the instanton is argued to be the one that joins the two spacetimes at their common limit of the IR brane and the black brane sent to the Poincar\'{e} horizon. It is observed that the contribution to the rate from the parts of the instanton where the radion EFT does not suffice is subleading, and this simply puts restrictions on the parameters of the theory. This is the approach taken in the first paper that showed how to set up the calculation~\cite{Creminelli:2001th}, and has been utilized in subsequent refinements. In the second approach, a putative configuration for the 5D instanton is assumed on physical grounds, even if it does not solve the 5D equations of motion~\cite{Agashe:2020lfz}. It is then argued that the correct configuration will only have a lower action, so this off-shell configuration gives a useful lower bound on the rate. Both these approaches stay in the limit of small backreaction in the IR. 

In this work we will stay in the regime where an estimate for the rate is possible, i.e. having a small backreaction in the IR, using radion EFT, and using the controlled part of the instanton to calculate the rate. In this limit, the problem can be mapped to a false vacuum decay problem in QFT at finite temperature. The dominant Euclidean action is assumed to be (Euclidean) time independent, and a $SO(3)$ ansatz is assumed for the action of the bubble interpolating between the false and true vacua. One can then minimize the bubble action to calculate the bubble profile and the on-shell action. 

There are a couple of points that need to be highlighted in this discussion, which distinguish this calculation from a standard false vacuum decay problem in QFT. The first distinction comes from the fact that the radion $\varphi$ is not a dynamical degree of freedom in the deconfined phase. Therefore, the bubble profile does not interpolate between one value of $\varphi$ to another. Rather, the dynamical degree of freedom in the deconfined phase is something entirely different, and the two have to match smoothly at the bubble wall. This aspect is crucial in the thick wall calculations, where this matching is used to identify the correct boundary conditions that the bubble profile must satisfy to make the minimization problem well posed~\cite{Agashe:2020lfz, Mishra:2023kiu}. The second point concerns the effect of gravity. Note that we have chosen the true vacuum to satisfy $V(\left<\varphi\right>) = 0$ which automatically makes the false vacuum have a positive vacuum energy and makes it inflate if the 4D gravity is present.  This affects the criteria for the phase transition to complete. Another effect of gravity is to modify the expression for the rate itself. However it turns out the effect of gravity in the calculation for the rate can be ignored in these calculations, even though the vacuum energy in the false vacuum is positive. To understand this we need to compare the typical bubble size with the Hubble radius in the false vacuum.
The typical size of the bubble (away from the extreme thin-wall regime) is $\sim 1/T$ while the Hubble radius is $\sim 1/T_c (M_4/T_c)$. For ignoring gravity we need the bubble radius to be much smaller than the Hubble radius, i.e. the nucleation temperature $T_n$ should satisfy
\begin{align}
    T_n/T_c \gg T_c/M_4\:,
    \label{eq:condition-to-ignore-gravity-for-Sb}
\end{align}
where $M_4$ is the 4D Planck scale. As we will see, this condition is easily satisfied for the parameters chosen here.

With these caveats in mind, one can show on general scaling grounds that the bounce action scales as~\cite{Creminelli:2001th, Agashe:2019lhy, Agashe:2020lfz, Mishra:2023kiu}
\begin{align}
    S_b  =  \frac{N^2}{\left|\kappa_\text{total}\right|^{3/4}} \, \widetilde{S}_b (T/T_c, \cdots)\:,
\end{align}
where $\kappa_\text{total}$ is the overall coefficient of the quartic of the radion potential (i.e. includes $\kappa_\theta$ if present), $N^2$ is a measure of the degrees of freedom, and $\widetilde{S}_b$ captures the dependence on the temperature and other parameters of the theory. Note that as $T/T_c \to 1$, $S_b \to \infty$ so that the rate is zero: this is expected since the phase transition only makes sense for temperatures below the critical temperature. When $T/T_c \lesssim 1$, the bubble facilitating the transition is big (compared to its wall region) and we are in the thin wall regime. When $T/T_c \ll 1$, the bubble is smaller, and we are in the thick wall regime. The scaling with $N$ and $\kappa$ is the same for both regimes, but the functional dependence on $T/T_c$ changes. For the thin wall case, the functional dependence is~\cite{Creminelli:2001th}
\begin{align}
    \widetilde{S}_b(T/T_c) \:\: \propto\:\:\frac{1}{T/T_c}\frac{1}{(1-(T/T_c)^4)^2} \:,
\end{align}
and becomes large as $T/T_c\to1$, as expected.

For the thick wall case, one has to calculate $\widetilde{S}_b$ numerically. We will not go into the details of this calculation, and refer the reader to refs.~\cite{Agashe:2020lfz, Mishra:2023kiu}, which discuss these steps in detail. We will simply borrow a numerically fitted functional form for $\widetilde{S}_b$ as a function of $T/T_c$, which works in the entire range (i.e. both thin wall and away from it). The functional form that works well for the choice of parameters we will work with is 
\begin{align}
    \log \widetilde{S}_b  = a_0 + a_1\,(\log(T_c/T))^{b_1} + a_2\,(\log(T_c/T))^{2b_2}\:,
    \label{eq:Sb-function-thick-wall}
\end{align}
where the constants $a_0, a_1, a_2, b_1, b_2$ are obtained by a numerical fit.

For the phase transition to complete, we need the rate in a Hubble volume to be order unity, which translates to the bounce action being smaller than $S_b^\text{max}$ given by
\begin{align}
    S_b^{\text{max}} = -\log \left(\frac{4\cpi^8}{9}\frac{T_c^4}{T^4}\frac{(M_5\ell)^6 T_c^4}{M_4^4}\right)\:,
    \label{eq:Sb-max}
\end{align}
where $M_4 = (M_5^3\ell)^{1/2}$ is the 4D Planck scale. $S_b^\text{max}$ is independent of $\kappa$ and has a mild log dependence on $T/T_c$.

With this discussion, it is straightforward to understand the effect of a non-zero $\theta$ on the rate for the phase transition. From previous sections we know that the effect of $\theta$ is to change the coefficient of the quartic $\kappa$, and reduce the critical temperature $T_c$. More explicitly, the rate for the phase transition is given as
\begin{align}
    \Gamma = T^4 \exp \left(-\frac{N^2}{\left|\kappa + \kappa_\theta\right|^{3/4}}\widetilde{S}_b(T/T_c)\right) = T^4 \exp \left(-\frac{N^2}{\left|\kappa + a\, \theta^2\right|^{3/4}}\widetilde{S}_b(T/T_c)\right)\:, 
\end{align}
where $a$ is a model-dependent number that scales as $1/N^2$. We see that the $\theta$ dependence is of the form $e^{-\left|\kappa + a \, \theta^2\right|^{-3/4}}$. While $a\,\theta^2$ is always positive, $\kappa$ can be either positive or negative, which translates to the rate increasing (if $\kappa > 0$) or decreasing (if $\kappa < 0$). Crucially, a small change in $\theta$ can change the rate strongly. The effect is smaller if $N$ is larger, but then the entire rate is suppressed anyway. Note that this discussion of $\theta$ dependence is only for $\Gamma$ (or equivalently $S_b$). For the phase transition to complete, one also has to look at the effect of $\theta$ on the Hubble in the false vacuum, which is captured by $S_b^\text{max}$. There are two choices here: either one adjusts the vacuum energy to be zero at the new minimum, i.e. $V(\left<\varphi\right>_\theta)$ is adjusted to zero, or it is not. For the first choice, the positive vacuum energy of the false vacuum changes, which in turn means that $S_b^\text{max}$ changes. This new $S_b^\text{max}$ has the same form as in eq.~\eqref{eq:Sb-max}, if the $T_c$ appearing there is understood to be $T_c(\theta)$. This is what we will assume in the discussions to follow (but our result can be adjusted in a straightforward manner for the second choice). For the second choice where $V(\left<\varphi\right>_\theta)$ is not adjusted to zero, there is no change in the vacuum energy of the false vacuum and therefore $S_b^\text{max}$ is unchanged. The fact that there is now a positive vacuum energy in the true vacuum has two effects: $1)$ it can change the estimate for $S_b$ and $2)$, the Universe after the phase transition ends in an inflating phase. For the estimate for $S_b$ to not be affected, one again needs that the bubble radius be smaller than the Hubble radius. Since the Hubble in the true vacuum is necessarily smaller than that in the false vacuum (i.e. the Hubble radius in the true vacuum is larger than that of the false vacuum), and we already saw that the bubble radius is too small compared to the Hubble radius for the false vacuum, it follows that the effect on $S_b$ can be ignored (see discussion near eq.~\eqref{eq:condition-to-ignore-gravity-for-Sb}). The fact that the Universe can end up in an inflating phase after the phase transition is often studied in the so called early dark-energy (EDE) scenario, which has interesting phenomenology and the potential to resolve some of the cosmological tensions. It will be interesting to revisit this aspect and attempt a realistic phenomenological study of this scenario.

With these caveats mentioned, let us look at the effect of $\theta$ on the three types $\mathbf{A, B, C}$ next.
For radion potentials of type $\mathbf{A}$, the effect is quite small if we do not want to destabilize the minimum, since $\kappa_\theta$ is bounded from above (see eq.~\eqref{eq:kappa-restriction-typeA}) by a small number (since $\epsilon\ll1$ and $v_\ir^2/M_5^3 \lesssim 1$). For type $\mathbf{B}$, which originally had $\kappa<0$, a non-zero (positive) $\kappa_\theta$ can make the overall $\kappa\to0$. When that happens, the action increases sharply and leads to a very small rate for the transition. For type $\textbf{C}$, since $\kappa$ originally is already positive, adding a positive $\kappa_\theta$ makes it larger and reduces the action, thereby increasing the rate. However here one cannot increase the rate arbitrarily because the overall $\kappa$ should not be larger than order unity, otherwise we have to include backreaction into the analysis. 

\begin{figure}
    \centering
    \includegraphics[width=0.485\linewidth]{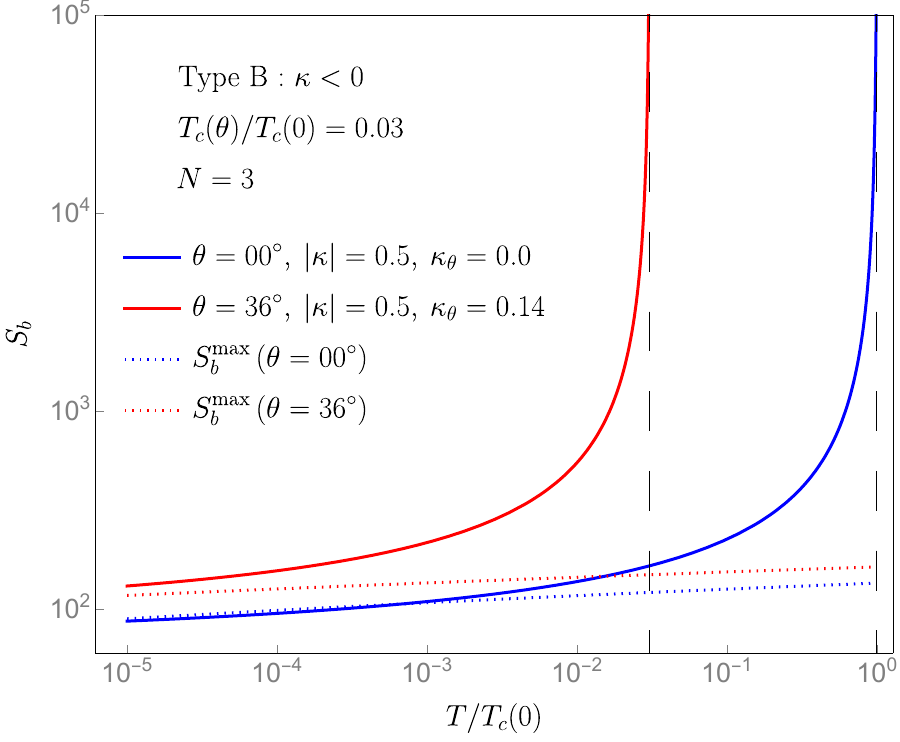}
    \:
    \includegraphics[width=0.485\linewidth]{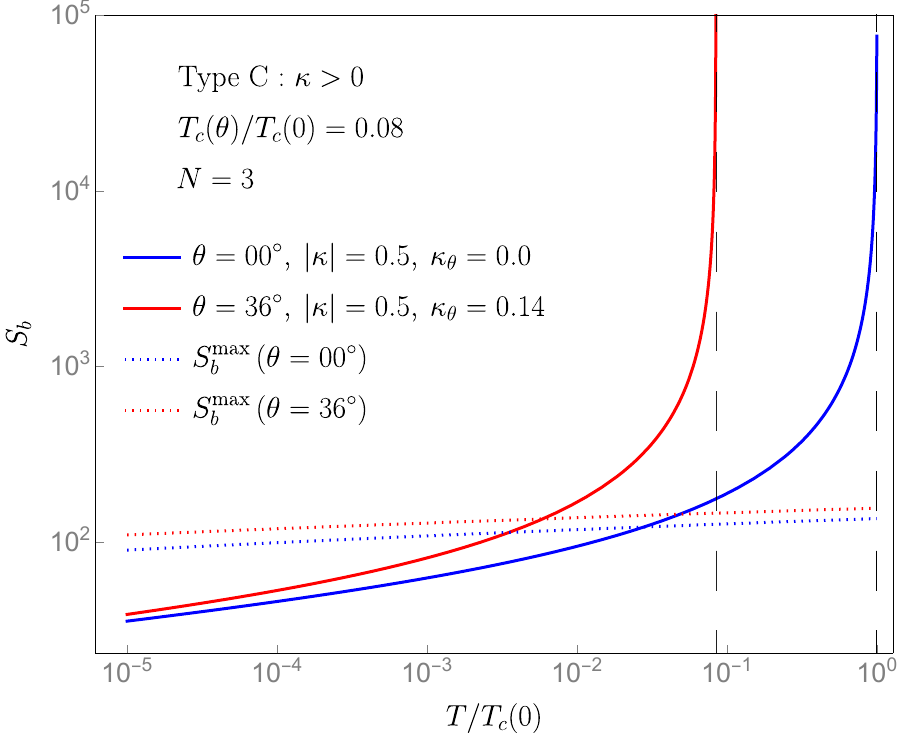}
    \caption{\small{The bounce action $S_b$ for $N=3$ as a function of $T/T_c(0)$, for $\theta = 0^\circ$ (blue solid) and $\theta = 36^\circ$ (red solid). Left panel shows for type $\mathbf{B}$ type radion potential where $\kappa < 0$ originally, and right panel shows for type $\mathbf{C}$ radion potential where $\kappa>0$ originally.  Dotted lines show $S_b^\text{max}$, which must be larger than the corresponding $S_b$ for the phase transition to proceed. For type $\mathbf{B(C)}$, we have taken $\kappa = -1/2(1/2), \epsilon = 1/10(-1/10)$ and $f_\sigma\ell = 0.62$ which gives $\kappa_\theta = 0.14$ for $N=3$ and $T_c(\theta)/T_c(0) = 0.03(0.08)$. For both these cases, the other parameters of the potential are chosen to have the minimum at $\theta = 0$, $\left<\varphi\right>\sim 10^{-15}$. We have used the functional form of $\widetilde{S}_b$ from eq.~\eqref{eq:Sb-function-thick-wall} which gives $a_0 = 146.3, a_1 = -154.2, a_2 = 11.3, b_1 = 0.02, b_2 = 0.07$ for type $\mathbf{B}$ and $a_0 = 199.0, a_1 = -195.7, a_2 = 0, b_1 = 0.005, b_2 = 1.515$ for type $\mathbf{C}$.}}
    \label{fig:BounceAction-typeBC}
\end{figure}

Figure~\ref{fig:BounceAction-typeBC} shows the bounce action $S_b$ for $N=3$ as a function of $T/T_c(0)$, for $\theta = 0^\circ$ (blue solid) and $\theta = 36^\circ$ (red solid). Left panel shows for type $\mathbf{B}$ radion potential where $\kappa < 0$ originally, and right panel shows for type $\mathbf{C}$ radion potential where $\kappa>0$ originally. Due to a non-zero $\theta$, there are two changes in the behavior of $S_b$. First, since the critical temperature reduces, the curve shifts to the left, and peaks at a value of $T/T_c(0) < 1$ (it is at $T/T_c(\theta) = 1$). Second, since a non-zero $\theta$ generates a $\kappa_\theta$, $S_b$ curve moves up or down depending on whether the overall coefficient of quartic increases or decreases in magnitude. For type $\mathbf{B}$ it increases, while for type $\mathbf{C}$ it decreases. Finally, dotted line shows $S_b^\text{max}$, which must be larger than $S_b$ for the phase transition to proceed. Since $S_b^\text{max}$ depends on $T_c$ logarithmically, and $T_c$ changes due to $\theta$, there are two $S_b^\text{max}$ curves, one for $\theta = 0$ (blue dotted) and one for $\theta\neq 0$ (red dotted). Note that since we have allowed $S_b^\text{max}$ to change from a non-zero $\theta$, we are assuming that the vacuum energy in the new minimum has again been adjusted to zero. If this was not assumed, $S_b^\text{max}$ would stay the same and in fig.~\ref{fig:BounceAction-typeBC} we should only look at the blue dotted line for $S_b^\text{max}$.

It is clear that there is a significant effect of having a non-zero $\theta$, specifically on when the phase transition happens. For example, for type $\mathbf{B}$ radion potential, without $\theta$ the phase transition would proceed around $T/T_c(0) \sim 10^{-3}$, whereas with a non-zero $\theta$ it does not happen till much lower temperatures $T/T_c(0) \sim 10^{-5}$, i.e. a factor of $100$ later. For type $\mathbf{C}$ radion potential, the phase transition at $\theta=0$ happens at $T/T_c(0) \sim 3\times10^{-2}$, but due to the presence of $\theta$ it happens around $T/T_c(0) \sim 6\times10^{-3}$ i.e. only a factor of $~10$ later. The difference in these two cases is because the behavior of $S_b$ at low temperatures, where it matters whether $\epsilon$ is positive or negative (i.e. whether the CFT deformation is getting larger or smaller), is different. This is evident by the fact that $S_b$ starts to saturate for type $\mathbf{B}$ but keeps reducing for type $\mathbf{C}$. We should note that given a $\theta$ and $N$, these curves are fixed. One can increase or decrease $S_b$ by changing $\theta$ or $N$, but it also changes $T_c(\theta)/T_c(0)$ accordingly, so there is not complete arbitrariness.

It is worth noting that due to the multi-branch nature, there is more than one minimum (in cases when they are not destabilized), to which the hot phase can finally end up in, post the transition. These minima are not degenerate. The theory is expected to settle in the minimum with the lowest free energy, after all the transients have decayed. We will have more to say about this in the next section.   

Note that here we have contented ourselves with just considering a quadratic bulk potential for $\Phi$, which is known to be constrained in letting the phase transition proceed and does not allow larger values of $N$. There are several well justified proposals~\cite{Hassanain:2007js, Dillon:2017ctw, vonHarling:2017yew, Baratella:2018pxi, Fujikura:2019oyi, Agashe:2019lhy, Agashe:2020lfz, Csaki:2023pwy, Mishra:2023kiu, Gherghetta:2025krk, Agrawal:2025wvf} on what kind of dynamics would relax this constraint, allowing for larger values of $N$, which is desirable for theoretical control. All of these still have the same $\kappa$ dependence of $S_b$, and having a non-zero $\theta$ will change the corresponding results in a similar way as here. The critical temperature will reduce, and the $S_b$ curve will move to the left. It will also move up or down depending on whether the overall coefficient of the quartic is increasing or decreasing in magnitude from having a non-zero $\theta$.

\section{Early Universe Implications}
\label{sec:EarlyUniverse}
We can apply the results from the previous sections to a strongly coupled confining dark sector undergoing a phase transition in the early universe. This dark sector has nothing to do with the scales in the SM so in principle the phase transition can proceed at very different temperatures. Such sectors have gathered interest in the literature due to the gravitational wave (GW) signals they can generate, and the observed GW signals by NANOGrav in the nanohertz frequency range~\cite{NANOGrav:2020bcs, NANOGrav:2023gor}. 

In the early universe, the temperature is time dependent and the specific functional form depends on the equation of state of the fluid that drives the expansion of the universe. Even though the calculations for the free energies of the deconfined and the confined phase and the rate for the phase transition are done at fixed temperature, one can apply them when the temperature is time-dependent. The implicit assumption is that the rate of change of temperature is slow and the dark sector is sufficiently strongly coupled, so the results from constant temperature are a good approximation. In the $\theta = 0$ case, this means that as the temperature redshifts in the early universe, the bounce action $S_b$ decreases till it crosses $S_b^\text{max}$, which is when the bubbles of true vacuum start to appear in the false vacuum, grow, merge and the phase transition proceeds. 

With a non-zero $\theta$, it is a natural possibility that $\theta$ is also time-dependent, i.e. $\theta \to \theta(t)$. The specific form of this time dependence is not fixed a priori. One can engineer a setup in which the angle is constant at early times, changes rapidly at some point to a lower value and then is constant again. For example, the corresponding axion field can be at some random value at high temperatures when there is no potential for it, and then evolve to its minimum as a potential for it is generated at intermediate temperatures (this is not the potential generated from the confinement itself, but is rather a UV contribution that was shielded at high temperatures). As another example, one can consider a complex scalar with a potential that admits multiple minima, and where the corresponding VEVs have a different phase. If this complex scalar couples to $\tr F\widetilde{F}$ by its phase, it can cause a change in $\theta$ and the complex scalar undergoes its transition. 

This change in $\theta$ can be gradual, or sharp, which depends on further details. It is more natural to assume that $\theta$ is order one and as the universe evolves, it becomes small. In that case, a very interesting scenario can occur in the early universe. We saw that with a non-zero $\theta$, for some choice of radion potential, the bounce action is enhanced enough to switch off the phase transition from proceeding till much lower temperatures. If $\theta$ evolves from an order one number to a small number, the phase transition can be switched off in the beginning and switched on suddenly. Effectively, this gives a mechanism to switch on the phase transition. Alternatively, this gives a way to generate some amount of supercooling, but in a controlled way. 

\begin{figure}
    \centering
    \includegraphics[width=0.95\linewidth]{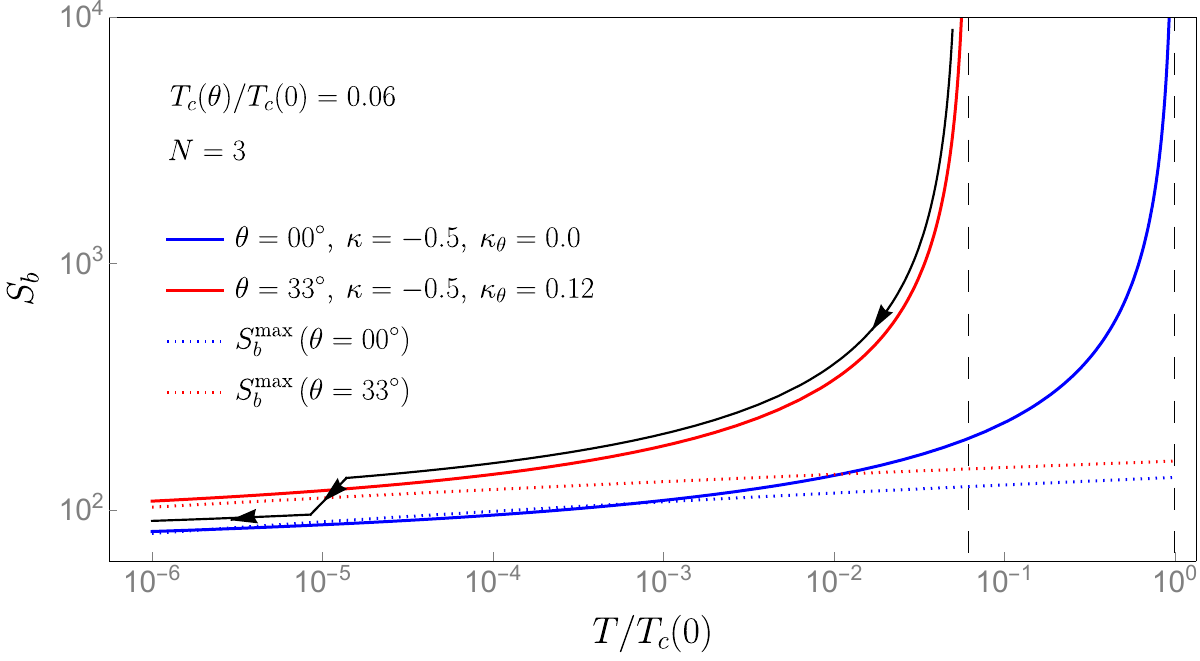}
    \caption{\small{A cartoon of an early universe scenario, where the angle $\theta$ transitions from $\sim 30^\circ$ at early times to $\sim 0^\circ$ at late times. At early times, the bounce action traces the red curve as a function of the red-shifting temperature and the action is too large for the phase transition to proceed. Once the $\theta$ transition happens, the bounce action starts to trace the blue curve, and is now low enough for the phase transition to proceed. For some choice of parameters, the change in $\theta$ can trigger the deconfinement to confinement phase transition.}}
    \label{fig:BounceAction-EarlyUniverseEvolution}
\end{figure}

One such scenario is shown in fig.~\ref{fig:BounceAction-EarlyUniverseEvolution}, where the time evolution is shown by the black lines with arrow. At early times when $\theta$ is non-zero, the bounce action traces the red line and is too large to make the phase transition possible. Once $\theta$ drops, it starts to trace the blue curve and now the phase transition can proceed easily. In fig.~\ref{fig:BounceAction-EarlyUniverseEvolution} we have shown the original $S_b^\text{max}$ at $\theta = 0$ and the changed $S_b^\text{max}$ at non-zero $\theta$. If we are in the scenario where the vacuum energy is again adjusted at the new minimum with a non-zero $\theta$, we should look at a new $S_b^\text{max}$, whereas if we do not do so, we should look at the original $S_b^\text{max}$. As we see, the general conclusion we want to draw, that the phase transition can be triggered by a change in $\theta$ is unaffected by this. 

Also note that when $\theta$ is dynamical, it presumably has a potential $U(\theta)$ which contributes to the vacuum energy. In the following schematic discussion we assume that the sector responsible for the time dependence of $\theta$ is a spectator, in the sense that its potential and kinetic energy are small compared to the false-vacuum energy driving the transition, $U_\theta(\theta)+\frac12 f_a^2 \dot\theta^2 \ll \rho_\text{vac}^\text{false} \sim 2\cpi^4 (M_5\ell)^3 T_c^4$, where $f_a$ is the decay constant of the dark axion. Since $(M_5\ell)^3\sim N^2$, it is plausible that this condition can be satisfied. Under this assumption the Hubble rate and hence $S_b^\text{max}$ are unaffected by the potential for $\theta$. If this condition is not satisfied, the axion and radion dynamics must be treated together, and the estimates will change. 

A possible consequence of such a dynamics is that the peak frequency of the emitted gravitational waves from bubble collisions, which in turn is decided by the parameter ``$\beta$'' (related to the slope of $S_b$), changes due to a non-zero $\theta$ (in the example in fig.~\ref{fig:BounceAction-EarlyUniverseEvolution}, it can be a factor of a few smaller). Further, the power in the gravitational waves from bubble collision, which scale as $\beta^{-2}$, also changes. Visibility of the GW signal would also depend on whether the Universe is inflating after the phase transition or not. It would be interesting to further investigate the model building possibilities in the scenario outlined here. 

We should note that even though the effect of $\theta$ is degenerate with a detuned IR brane tension as far as the radion potential is concerned (but only for one sign of detuning), they are not on the same footing if thought of as a trigger for the deconfinement to confinement phase transition. While it makes sense for $\theta$, a UV quantity, to change while the theory is in the deconfined phase, it is not clear what changing the detuning of the IR brane tension would mean while the theory is in the deconfined phase, i.e. when the IR brane has not even formed. 

We would like to mention an interesting scenario that can unfold in this setup. As briefly mentioned in the previous section, due to the multi-branch nature, there is more than one minimum to which the hot phase can transition to. These minima are however not degenerate. Due to the small rate of phase transition and the fact that there is always some amount of supercooling, it is likely that at a temperature when the phase transition proceeds, multiple minima will be accessible as an option to tunnel to. This will give rise to domain walls in the early universe. Since the minima are not degenerate, these domain walls are not stable, and the regions with higher free energy will shrink. Ultimately, the domain walls will collapse and annihilate. In the process, they will release gravitational waves. Since the gravitational wave signatures of domain wall collapse are different from bubble collision, it would leave a distinct signal in the experiments than if the transition was simply from bubble collisions. We will leave a more detailed phenomenological investigation of this scenario for the future.

Finally, as briefly alluded to in the earlier section, there is a domain wall solution at $\theta = \pi$ whose bulk $\sigma$ profile can be constructed in the 5D model. It would be interesting to study the dynamics of such a domain wall, and in particular, what early Universe signals can arise from the generation and evolution of such an object in a dark sector.

\section{Beyond the Small Backreaction Limit}
\label{sec:Generalizations}
In this work we have stayed in the limit of small backreaction in the IR, which is justified when the radion is light compared to the KK scale. This allowed the free energy and the rate for the phase transition to be computed in a somewhat controlled way. Being in the small backreaction limit is an idealization at best---in 10D examples dual to confining field theories, the geometry deviates strongly from a pure AdS background as one gets closer to the place where the geometry caps off, i.e. the location of the IR brane in the 5D simplified picture. When the backreaction is included, there are new features that appear in the phase diagram, most notably in the deconfined phase. There is a new scale that appears, a \textit{minimum} temperature $T_\text{min}$~\cite{Mishra:2024ehr} which is close to but smaller than $T_c$, below which the black brane phase does not exist. Information about this scale is absent in the black brane phase without including backreaction. In fact this aspect of the phase diagram is very similar to that of black holes in global AdS, which also have a minimum temperature and a critical temperature (for the Hawking-Page transition)~\cite{Hawking:1982dh}, the minimum temperature is slightly below the critical temperature, and the scale is provided by the radius of the sphere on which the dual CFT lives.

Close to the minimum temperature, the black brane phase has a classical instability, as seen by the speed of sound becoming purely imaginary. It is currently an open problem to calculate how this instability plays out in the dynamics of the phase transition. The generic expectation is that as the temperature cools below the minimum temperature, this classical instability will drive the theory to the confined phase, since that definitely exists even at much lower temperatures. It is important for these discussions that the minimum temperature is slightly smaller than the critical temperature, which makes sense physically---any process that drives the system to change to the confined phase should only switch on after the critical temperature. This is certainly seen in the UV examples as well as in the AdS global case. 

In the presence of a non-zero $\theta$, we run into a puzzle. First note that the contribution to the free energy on the deconfined side, even in the case of strong backreaction in the IR, is still zero. This can be argued as follows. When the backreaction is large, the equations of motion for $\Phi$ and the 5D metric mix with each other and strongly affect each other. The equation of motion for $\sigma$ is still the same as before
\begin{align}
    \partial_z\left(\sqrt{g} g^{55} \partial_z\sigma \right) = 0\:,
\end{align}
except that the metric is not purely AdS any more in the IR. Even though that is the case, we expect $g^{55} \to 0$ as $z\to z_h$. Let us denote the $z$ dependent term $\sqrt{g} g^{55}$ as $A(z)(1 - z/z_h)$, making the zero at $z= z_h$ explicit: $A(z)$ is a regular non-zero function of $z$ near the horizon (e.g. for the pure AdS case, $A(z) = (\ell^3/z^3) (1+z/z_h)(1+z^2/z_h^2)$). Close to the horizon one can solve for $\sigma$:
\begin{align}
    \sigma(z) = \sigma_0 + \int \dd z\, \frac{1}{A(z)(1 - z/z_h)}\:\sim \sigma_0 - \frac{z_h}{A(z_h)}\log(1-z/z_h)\:, 
\end{align}
where $\sigma_0$ is a constant. We see that regularity at the horizon will again require the coefficient of the log term to vanish and give a constant $\sigma$. One can alternatively argue that since $\sigma = \sigma_0$ is a solution to the equations of motion and is regular at the horizon, it is the unique solution (the constant will be set by the UV boundary condition). In this discussion we have ignored the bulk potential generated for $\sigma$ from instanton effects, which is expected to be suppressed as $e^{-N}$. With this caveat, as in the small backreaction case, it is clear that there will be no contribution to the free energy of the deconfined phase from the $\sigma$ action, the topological susceptibility in the deconfined phase vanishes, and the minimum
temperature is independent of $\theta$. However we saw earlier that the critical temperature decreases with $\theta$. While we derived this in the small backreaction limit, this is expected to hold even at large backreaction. Certainly on the lattice where the same behavior is seen, one is not in the small backreaction limit.

This is the puzzle: while $T_c$ decreases with $\theta$, $T_\text{min}$ does not. Naively it seems that one can have a scenario where $T_\text{min} > T_c$. It is not clear whether this is a pathological behavior and if so, what is the resolution. It is likely that a computation of these quantities at strong backreaction would resolve the issue. In fact, the $\theta$ dependence of the vacuum energy in the type IIA model does not keep growing quadratically with $\theta$ for larger values of $\theta$, but rather saturates~\cite{Dubovsky:2011tu}. This suggests that the critical temperature will also saturate for larger values of $\theta$. Nonetheless it is not a guarantee that the critical temperature will stay above the minimum temperature. At the very least, this requirement will impose non-trivial constraints on the bulk potential for $\Phi$ in our setup, or equivalently, on the beta function of the scalar deformations of the CFT. We will simply pose this as a puzzle here and defer a more careful analysis to future.

\section{Conclusion and Outlook}
\label{sec:Conclusion}
We have presented here a simple five-dimensional model to study properties of a strongly coupled holographic gauge theory with a $\theta$ term. The construction was motivated by looking at a class of UV complete models. We showed that a free scalar in the bulk, but with appropriate boundary conditions on the UV and IR boundaries, can capture the effect of $\theta$ on the dynamics. The boundary conditions are motivated by looking at higher dimensional examples where $\theta$ comes from a gauge field winding around a compact spatial dimension which shrinks to zero size in the IR. The IR boundary condition, that the field that is sourced by $\theta$ should vanish on the IR brane, is further justified because an angular variable must vanish at the origin of coordinates. Taking the dual 5D scalar field and the boundary conditions as the starting point, we applied this bottom up model to the question of deconfinement to confinement phase transition. We showed that due to a non-zero $\theta$, the critical temperature for the phase transition reduces quadratically with $\theta$ (to leading order), matching the lattice expectation. We also showed in the process how the vanishing of the topological susceptibility in the deconfined phase comes about, explaining the drop in topological susceptibility across the critical temperature, as seen on the lattice. We then used the setup to calculate the effect of $\theta$ on the rate of the phase transition, showing that the bounce action responsible for the transition can change significantly, thereby changing details of the phase transition. We sketched out how this can play out in the early universe, and has the potential to give a controlled way to generate supercooling. In all these discussions we stayed in the limit of small backreaction in the IR. 

There are several open directions that are worth mentioning. It is natural to consider relaxing the assumption of small backreaction in the IR. As discussed in section~\ref{sec:Generalizations}, there is a puzzle in this limit where naively it seems that the critical temperature of the deconfined phase, which is usually larger than the minimum temperature, can become smaller in the presence of $\theta$.  Requiring the critical temperature to be larger will impose non-trivial constraints on the bulk potential for $\Phi$ in our setup, or equivalently, on the beta function of the scalar deformations of the CFT. Such constraints on allowed scalar potentials in a theory of quantum gravity are active research directions, and connecting the present discussion to the existing proposals can yield further insight. It would be instructive to write down an explicit model and further investigate this.

In the setup we considered, the bulk degrees of freedom are the graviton and the two scalars. It is straightforward to include additional matter in the bulk, make $\theta$ dynamical by adding the corresponding axion and derive how the low energy interactions among the light fields emerge. It would be interesting to see if this has any effect on the dynamics of phase transition. 

Two of the results presented here would be useful to understand from a dual 4D point of view. For example, it is intriguing that the confining minimum is lost for some values of $\theta$, at least for some types of stabilizing dynamics. Understanding this effect from a dual point of view can tell whether this is a generic expectation. We also saw that there is no contribution to the free energy of the deconfined phase from a non-zero $\theta$, i.e. the topological susceptibility  vanishes. It would be insightful to understand this more generally, purely by field theory arguments. 

The model building aspects in the early universe scenario are worth further consideration. It would be useful to construct an explicit model where a changing $\theta$ generates a given amount of supercooling, understand how much supercooling is possible for reasonable models and parameters, and do a more careful analysis of the effect of such a dynamics on the gravitational wave signals and their experimental sensitivity. It would also be interesting to investigate the role played by the multi-branch structure which gives more than one minimum as an option to tunnel to, specifically whether the transition happens in one go or in steps, when domain walls form and annihilate, and what kind of a cascaded GW signal this can generate.

Finally, it would be useful to explicitly construct the domain wall at $\theta = \cpi$, putting in the localized degrees of freedom and study properties of Wilson lines or confining strings near this wall. The 5D uplift of such a domain wall can provide a useful model to understand some of the properties of $SU(N)$ YM at $\theta = \cpi$ when there is a holographic description available.

\vspace{5mm}
\noindent
\textbf{Note added:} During the completion of this work, I was made aware of a related work~\cite{Csaki:2026arv} that similarly argues for the IR boundary condition advocated in the current work.

\section*{Acknowledgments}
I would like to thank Matt Reece for useful discussions and comments on the draft. I also thank Nathaniel Craig and Lisa Randall for useful comments on the draft. I acknowledge support from the \textit{Gravity Spacetime and Particle Physics} (GRASP) initiative at Harvard University. I thank the authors of~\cite{Csaki:2026arv, Csaki:2026qjl} for sharing their unpublished work with me. I also thank the anonymous referee for several useful suggestions and specifically for asking to expand discussions related to the structure of the potential at $\theta = \cpi$, spontaneous CP breaking, and the dynamical object that interpolates between them.

\bibliographystyle{utphys}
\bibliography{references}

\end{document}